\documentclass[epj]{svjour}

\usepackage[autostyle,italian=guillemets]{csquotes}
\usepackage[utf8]{inputenc}
\usepackage[nowrite,infront,standard]{frontespizio}
\usepackage[english]{babel}
\usepackage{physics}
\usepackage{amsmath,amssymb}
\usepackage [svgnames]{xcolor}
\usepackage{xcolor}

\usepackage{mathrsfs}
\usepackage{subfloat}
\usepackage{frontespizio}
\usepackage{chemmacros}
\usepackage{float}
\usepackage{graphicx} 
\usepackage{bm}
\usepackage{appendix}
\usepackage{hyperref}
\usepackage{midpage}
\usepackage{placeins}
\usepackage{subfig}

\begin{document}
\title{Quantifying quantumness in three-flavor  neutrino oscillations}
\author{ V.~A.~S.~V.~Bittencourt \inst{1}, M.~Blasone\inst{2,3}, S.~De Siena\inst{4} and C.~Matrella\inst{2,3}
}                     
%
%
\institute{ISIS (UMR 7006), Universit\'{e} de Strasbourg, 67000 Strasbourg, France \and Dipartimento di Fisica, Universit\`a degli Studi di Salerno, Via Giovanni Paolo II, 132 84084 Fisciano, Italy \and INFN, Sezione di Napoli, Gruppo Collegato di Salerno, Italy \and Universit\`a degli Studi di Salerno (Ret. Prof.),  email: silvio.desiena@gmail.com }
\date{Received: date / Revised version: date}
%
\abstract{We characterize quantum correlations encoded in a three-flavor oscillating neutrino system by using both plane-wave and wave-packet approach. By means of the Complete Complementarity Relations (CCR) we study the trade-off  of predictability, local coherence and non--local correlations  in terms of the relevant parameters, chosen from recent neutrino experiments. Although the CCR describe very well the  contributions  associated to bipartite correlations, an attempt of promoting these relations to include the genuine tri-partite contributions in the pure-state case leads to a  not completely meaningful result. However, we provide an analysis of the genuine tripartite contributions  both for the  pure instance and for the mixed case, independently of CCR.
\PACS{
      {PACS-key}{discribing text of that key}   \and
      {PACS-key}{discribing text of that key}
     } 
} 
\authorrunning{V. Bittencourt, M.~Blasone, S.~De~Siena and C.~Matrella}
\maketitle
\section{Introduction}
\label{intro}

In the last few years, elementary particles as neutrinos have been investigated in the context of quantum information \cite{Blas1}--\cite{EURnoi}. The property of neutrinos  to interact very weakly  and to deeply penetrate  into matter makes these particles  interesting candidates for applications of quantum information beyond photons. In fact, classical communication using a neutrino beam was demonstrated in \cite{Stancil}. The characterization of quantum correlations in such systems is therefore important for the development of algorithms and protocols that can harness not only quantum entanglement but also  other resources, such as quantum discord (QD) \cite{QD} and coherence \cite{Braumgratz}.
 

Specifically, the quantum nature of neutrino oscillations has been studied in terms of  entanglement \cite{Blas1}--\cite{Banarjee}, Bell and Leggett-Garg inequalities \cite{Gango}--\cite{chatto}  and in terms of other various aspects such as quantum coherence, steering \cite{Song,Uola}, coherence in conncection with mixedness \cite{Dixit}, non-local advantage of quantum coherence (NAQC) \cite{Song1}--\cite{Us}. Furthemore, neutrino oscillations have  been also analyzed in terms of entropic uncertainty relations  \cite{Wang1,EURnoi}.

Above aspects are also an interesting perspective to study more fundamental features of neutrino physics. In fact, entanglement and quantum correlations go back to the deepest nature of quantum physics and  of the fundamental interactions, and can be connected to  properties and symmetries exploited in  particle physics \cite{Cerveira}--\cite{Afik:2022dgh}.

Neutrino states exhibit a complex correlation structure:
 flavor oscillations directly affect the informational content shared between different neutrino flavors, and between flavor and other degrees-of-freedom, such as spin \cite{Gennaro}.
In this framework, complete complementarity relations (CCR) provide an effective way to characterize quantum correlations in bi- and multi-partite systems \cite{Basso2020} and, in particular, can be applied to the description of quantum correlations intrinsic to neutrino systems \cite{CCR2,CCR3piane}. 
The concept of complementarity is often associated with wave-particle duality \cite{Bohr}: it is summarized in the statement that a quantum system may possess properties which are equally real but mutually exclusive, in the sense that the more information one has about one aspect of the system, the less information can be obtained about the other. In the context of the two-slit experiment, CCR  can be formalized \cite{Englert,Wootters} by defining a predictability $P$, associated to the knowledge of the path of the particle, and a visibility $V$, connected with the capacity to distinguish the interference fringes:
\begin{equation}
P^{2}+V^{2}\le 1.
\label{1}
\end{equation}
Complementarity relations as in Eq.(\ref{1}) are saturated only for pure single-partite quantum states. In \cite{Jakob2010} it is shown that for a bipartite state we have to consider a third entry $C$ -- representing the correlation between the subsystems -- in order to obtain a complete complementarity relation:
\begin{equation}
P_k^{2}+V_k^{2} +C^2 =  1,\qquad k=1,2.
\label{2}
\end{equation}
The quantities associated with the wave-particle duality generate local, single-partite realities, while $C$ generates an exclusive bipartite nonlocal reality.

While CCR have been used to describe the interplay between the different correlations for two flavor mixing and for plane-wave three-flavor mixing, a complete picture of neutrino correlations in the tri-partite mixed state case induced by wave-packet dynamics, a situation phenomenologically appealing, is still lacking.


In this paper, we complete the CCR analysis of neutrino oscillations by considering the wave packet instance (mixed state) for three flavor mixing, and show that a rich structure of correlations emerges among the various bi-partitions of the system. For the pure state case, by exploiting polygamy relations, we attempt to include both bipartite and tripartite contributions to the CCR. To this aim, we include in the CCR the so-called residual correlation \cite{Residual,Residual2}, but we are not able to  identify this term  with the genuine tripartite correlation because the three possible expressions of residual correlation do not coincide among themselves. However, the genuine tripartite correlation can be quantified both as the average of the three residual correlations associated to the three single-partite subsystems and as the average of the three bipartite correlations, which provide the same result although the individual contributions are different. Moreover, a similar analysis has been extended to mixed states.

The paper is organized as follows. In  Section~2 we briefly review the formalism of CCR for pure  states, in which suitable bipartite terms are taken into consideration.   In Section~3, we consider the CCR for a mixed tri-partite state and discuss the results for a three-flavor neutrino system, by using a wave packet approach. Tripartite correlations for the case of pure and mixed states are discussed in  Section~4.  Finally, Section 5 is devoted to conclusions and outlook. In appendix \ref{A} some basics of neutrino oscillation theory are presented. Appendix \ref{B} contains some explicit expressions for the quantities of interest.

\section{Complete complementarity relations for pure systems}
\label{sec:1}
CCR have recently attracted much attention because they represent useful tool capable to encompass,  in quantum systems, various  correlations and their interplay. In subsections \ref{sub2.1} and \ref{sub2.2} we briefly review the formalism of CCR for bi- and tri-partite pure states, which we will subsequently use to describe quantum correlations in neutrino flavor oscillations.

\subsection{CCR for bipartite states}
\label{sub2.1}
Let us consider the general framework of \cite{Basso2020}: a bipartite state is represented as a vector in the Hilbert state $\mathcal{H}_{A}\otimes \mathcal{H}_{B}$ of dimension $d=d_{A}d_{B}$, where $d_{A}$ and $d_{B}$ are the dimension of the subsystem A and B, respectively. The set of  tensor products  $\{\ket{i}_{A}\otimes \ket{j}_{B}=\ket{i,j}_{AB}\}_{i,j=0}^{d_{A}-1,d_{B}-1}$ represents an orthonormal basis for $\mathcal{H}_{A}\otimes \mathcal{H}_{B}$, where $\{\ket{i}_{A}\}_{i=0}^{d_{A}-1}$ and $\{\ket{j}_{B}\}_{j=0}^{d_{B}-1}$ are  the local bases for the spaces $\mathcal{H}_{A}$ and $\mathcal{H}_{B}$, respectively. In this basis, the density matrix of any bipartite state is: 
\begin{equation}
\rho_{A,B}=\sum_{i,k=0}^{d_{A}-1} \sum_{j,l=0}^{d_{B}-1}\rho_{ij,kl}\ket{i,j}_{AB}\bra{k,l}.
\label{3}
\end{equation}
 The state of subsystem A(B) is obtained by tracing over B(A). For example, for subsystem A, we have:
\begin{equation}
\rho_{A}=\sum_{i,k=0}^{d_{A}-1} \left( \sum_{j=0}^{d_{B}-1} \rho_{ij,kj} \right)\ket{i}_{A}\bra{k} \equiv \sum_{i,k=0}^{d_{A}-1} \rho_{ik}^{A}\ket{i}_{A}\bra{k},
\label{4}
\end{equation}
with a similar form for the subsystem B.

In general, even if the joint state $\rho_{A,B}$ is pure, the states of the subsystems A and B are not, which implies that some information is missing when the state of a single subsystem is considered. 
The missing information is shared via correlations with the subsystem B \cite{Qian}. Such an interplay between correlations encoded in subsystems of a pure bipartite system, and correlations shared among them, can be described via the CCR which takes the form
\begin{equation}
P_{\rm{hs}}(\rho_{A})+C_{\rm{hs}}(\rho_{A})+C_{\rm{hs}}^{nl}(\rho_{A|B})=\frac{d_{A}-1}{d_{A}},
\label{5}
\end{equation}
where $P_{\rm{hs}}(\rho_{A})=\sum_{i=0}^{d_{A}-1}(\rho_{ii}^{A})^{2}-\frac{1}{d_{A}}$ is the predictability measure, $C_{\rm{hs}}(\rho_{A})=\sum_{i\ne k}^{d_{A}-1}|\rho_{ik}^{A}|^{2}$ is the Hilbert-Schmidt quantum coherence (a measure of visibility), and $C_{\rm{hs}}^{nl}(\rho_{A|B})=\sum_{i\ne k, j\ne l}|\rho_{ij,kl}|^{2}-2\sum_{i\ne k, j<l}\Re (\rho_{ij,kj} \rho_{il,kl}^{*})$ is called the non local quantum coherence, that is, the coherence shared between A and B.

Another form of CCR can be obtained by defining the predictability and the coherence measures in terms of the von Neumann entropy:
\begin{equation}
C_{{\rm{re}}}(\rho_{A})+P_{{\rm{vn}}}(\rho_{A})+S_{{\rm{vn}}}(\rho_{A})=\log_2 d_{A},
\label{6}
\end{equation}
where $C_{{\rm{re}}}(\rho_{A})=S_{{\rm{vn}}}(\rho_{A,\, {\rm{diag}}})-S_{{\rm{vn}}}(\rho_{A})$ is the relative entropy of coherence, $S_{{\rm{vn}}}(\rho)$ denotes the von Neumann entropy of  $\rho$ and $\rho_{A,\, {\rm{diag}}}=\sum_{i=1}^{d_{A}} \rho_{ii}^{A}\ket{i}\bra{i}$. $P_{{\rm{vn}}}(\rho_{A})\equiv\log_2 d_{A}-S_{{\rm{vn}}}(\rho_{A,\, {\rm{diag}}})$, is a measure of predictability. 
Eqs.(\ref{5},\ref{6}) can be exploited in different situations, but are not completely equivalent, as we will see in the following when we consider the problem of the genuine tripartite contribution.

\subsection{CCR for  tripartite states}
\label{sub2.2}
In \cite{Basso2020} the generalization of the CCR for tri-partite pure states is obtained.
Let us consider a tri-partite state represented as a vector in the Hilbert state $\mathcal{H}_{A}\otimes \mathcal{H}_{B}\otimes\mathcal{H}_{C}$ of dimension $d=d_{A}d_{B}d_{C}$, where $d_{A}$, $d_{B}$, $d_{C}$ are the dimension of the subsystem A, B and C, respectively.  The set of the tensor products $\{\ket{i}_{A}\otimes \ket{j}_{B}\otimes \ket{k}_{C}=\ket{i,j,k}_{ABC}\}_{i,j,k=0}^{d_{A}-1,d_{B}-1,d_{C}-1}$ represents an orthonormal bases for $\mathcal{H}_{A}\otimes \mathcal{H}_{B}\otimes\mathcal{H}_{C}$, where $\{\ket{i}_{A}\}_{i=0}^{d_{A}-1}$, $\{\ket{j}_{B}\}_{j=0}^{d_{B}-1}$ and $\{\ket{k}_{C}\}_{k=0}^{d_{C}-1}$ are  the local basis for the spaces $\mathcal{H}_{A}$, $\mathcal{H}_{B}$ and $\mathcal{H}_{C}$, respectively. In this basis, the density matrix of any tri-partite state is: 
\begin{equation}
\rho_{A,B,C}=\sum_{i,l=0}^{d_{A}-1} \sum_{j,m=0}^{d_{B}-1}\sum_{k,n=0}^{d_{C}-1}\rho_{ijk,lmn}\ket{i,j,k}_{ABC}\bra{l,m,n}.
\label{7}
\end{equation}
 The state of subsystem A is obtained by tracing over B and C:
\begin{equation}
\rho_{A}=\sum_{i,l=0}^{d_{A}-1} \left( \sum_{j=0}^{d_{B}-1}\sum_{k=0}^{d_{C}-1} \rho_{ijk,ljk} \right)\ket{i}_{A}\bra{l} \equiv \sum_{i,l=0}^{d_{A}-1} \rho_{il}^{A}\ket{i}_{A}\bra{l},
\label{8}
\end{equation}
with a similar form for the subsystems B and C.

The complete complementarity relation to consider for subsystem A is:
\begin{equation}
P_{\rm{hs}}(\rho_{A})+C_{\rm{hs}}(\rho_{A})+C_{\rm{hs}}^{nl}(\rho_{A|BC})=\frac{d_{A}-1}{d_{A}},
\label{9}
\end{equation}
where the non local coherence is given by:
\begin{equation}
\begin{split}
C_{\rm{hs}}^{nl}(\rho_{A|BC})=&\sum_{i\ne l}\left( \sum_{ j\ne m, k\ne n}+
\sum_{ j=m, k\ne n}+ \sum_{ j\ne m, k= n}\right)|\rho_{ijk,lmn}|^{2}\\
&-2\sum_{i\ne l}\left( \sum_{ j=m, k<n}+ \sum_{ j<m, k= n}+ \sum_{ j<m,k\ne n}\right)\Re (\rho_{ijk,ljk} \rho_{imn,lmn}^{*}).
\end{split}
\label{10}
\end{equation}

The other form of the CCR, Eq.(\ref{6}), is still valid for the single-partite subsystems A, B and C. But an interesting behaviour comes out when we consider the three possible bipartite subsystems AB, AC and BC.  For subsystem AB the following CCR is valid
\begin{equation}
C_{{\rm{re}}}(\rho_{AB})+P_{{\rm{vn}}}(\rho_{AB})+S_{{\rm{vn}}}(\rho_{AB})=\log_2 (d_{A}d_{B}),
\label{11}
\end{equation}
where, in contrast to the local coherence of a single-partite subsystem, the local coherences for bipartite subsystems do not vanish.
For the first time we explore such an effect for a neutrino system in the next sections.


\subsection{CCR for pure neutrino states}

In \cite{CCR2} we have analyzed the CCR for a bipartite neutrino state. 
We briefly recall the principal results. Let us consider an initial electronic neutrino:

\begin{equation}
\begin{aligned}
\ket{\nu_{e}(t)} &=a_{ee}(t)\ket{e}+a_{e\mu}(t)\ket{\mu} \\
&=a_{ee}(t)\ket{10}+a_{e\mu}(t)\ket{01}.
\end{aligned}
\label{12}
\end{equation}
From now on, we use the correspondence between flavor states and multi-qubit states established in \cite{Blas1}. For the case of two-flavor mixing, each flavor state corresponds to a two-qubit state describing the occupancy of a given flavor mode, thus 
\begin{equation}
    \vert \nu_e \rangle \equiv \vert 1\rangle_e \otimes \vert 0 \rangle _\mu \equiv \vert 1  0 \rangle
    \label{qubit},
\end{equation} and analogously for the three-flavor case.
By constructing the density matrix for the state $\rho_{A,B}$ and by tracing to obtain the density matrices for the subsystems $\rho_{A}$ and $\rho_{B}$, it is simple to check that  Eq.(\ref{5}) is verified. In fact,
$P_{\rm{hs}}(\rho_{A})=P_{ee}^{2}+P_{e\mu}^{2}-\frac{1}{2}$, $C_{\rm{hs}}(\rho_{A})=0$ and $C_{\rm{hs}}^{nl}(\rho_{AB})=2P_{ee}P_{e\mu}$, where we use $|a_{ee}(t)|^{2}=P_{ee}$, $|a_{e\mu}(t)|^{2}=P_{e\mu}$ and $P_{ee}+P_{e\mu}=1$.  Furthermore, it is simple to see that $\rho_{A}=\rho_{A,\, {\rm{diag}}}$ and, consequently, $S_{{\rm{vn}}}(\rho_{A})=S_{{\rm{vn}}}(\rho_{A,\, {\rm{diag}}})$. As result, $C_{{\rm{re}}}(\rho_{A})=0$, $P_{{\rm{vn}}}(\rho_{A})=|a_{ee}|^{2}\log_{2}|a_{ee}|^{2}+|a_{e\mu}|^{2}\log_{2}|a_{e\mu}|^{2}$ and $S_{{\rm{vn}}}(\rho_{A})=-|a_{ee}|^{2}\log_{2}|a_{ee}|^{2}-|a_{e\mu}|^{2}\log_{2}|a_{e\mu}|^{2}$. Since the dimension of subsystem A is $d_{A}=2$,  $\log_{2}d_{A}=1$ and Eq.(\ref{6}) is satisfied.

Although in the case of a bipartite pure neutrino   state, for both Eqs.(\ref{5}) and (\ref{6}), the local coherence term is zero,
it is natural to ask what happens in the case of a  tripartite neutrino state, in which there are bipartite subsystems with their own specific internal structure. We will see indeed that in this case the local coherence terms are non vanishing and  depend on the chosen bipartition.

For three flavor mixing,  exploiting the correspondence in Eq.(\ref{qubit}), the time evolution of a neutrino state initially ($t=0$)  in a flavor $\alpha = e, \mu, \tau$,  reads
\begin{equation}
\ket{\nu_{\alpha}(t)}=a_{\alpha e}(t)\ket{100}+a_{\alpha\mu}(t)\ket{010}+a_{\alpha\tau}(t)\ket{001},
\label{13}
\end{equation}
The amplitudes $a_{\alpha e, \mu, \tau}(t)$ depend on neutrino mixing angles and mass differences.

The density matrix associated to this state is given by:

\begin{equation}
\rho_{e\mu\tau}^{\alpha}=
\begin{pmatrix}
0&0&0&0&0&0&0&0\\
0&\rho_{22}^{\alpha}&\rho_{23}^{\alpha}&0&\rho_{25}^{\alpha}&0&0&0\\
0&\rho_{32}^{\alpha}&\rho_{33}^{\alpha}&0&\rho_{35}^{\alpha}&0&0&0\\
0&0&0&0&0&0&0&0\\
0&\rho_{52}^{\alpha}&\rho_{53}^{\alpha}&0&\rho_{55}^{\alpha}&0&0&0\\
0&0&0&0&0&0&0&0\\
0&0&0&0&0&0&0&0\\
0&0&0&0&0&0&0&0
\end{pmatrix}
\label{14}
\end{equation}
where the matrix elements are written as:
\begin{eqnarray}
\rho_{22}^{\alpha}=&|a_{\alpha\tau}(t)|^{2};& \hspace{1cm}\rho_{23}^{\alpha}=\rho_{32}^{\alpha*}=a_{\alpha\tau}(t)a^{*}_{\alpha\mu}(t);
\hspace{1cm}\rho_{25}^{\alpha}=\rho_{52}^{\alpha*}=a_{\alpha\tau}(t)a^{*}_{\alpha e}(t);\\
\rho_{33}^{\alpha}=&|a_{\alpha \mu}(t)|^{2};&\hspace{1cm}\rho_{35}^{\alpha}=\rho_{53}^{\alpha*}=a_{\alpha\mu}(t)a^{*}_{\alpha e}(t);
\hspace{1cm}\rho_{55}^{\alpha }=|a_{\alpha e}(t)|^{2}.
\end{eqnarray}

The corresponding oscillation probabilities are $P_{\alpha e}(t)=|a_{\alpha e}(t)|^{2}$, $P_{\alpha \mu}(t)=|a_{\alpha\mu}(t)|^{2}$,
 $P_{\alpha \tau}(t)=|a_{\alpha\tau}(t)|^{2}$. By tracing with respect one of the subsystems we can obtain the reduced density matrix for bipartite subsystems $e\mu$, $e\tau$, $\mu\tau$, which are, respectively:
\begin{equation}
\rho^{\alpha}_{e\mu}=\begin{pmatrix}
\rho_{22}^{\alpha}&0&0&0\\
0&\rho_{33}^{\alpha}&\rho_{35}^{\alpha}&0\\
0&\rho_{53}^{\alpha}&\rho_{55}^{\alpha}&0\\
0&0&0&0
\end{pmatrix},
\hspace{0.4cm}
\rho^{\alpha}_{e\tau}=\begin{pmatrix}
\rho_{33}^{\alpha}&0&0&0\\
0&\rho_{22}^{\alpha}&\rho_{25}^{\alpha}&0\\
0&\rho_{52}^{\alpha}&\rho_{55}^{\alpha}&0\\
0&0&0&0
\end{pmatrix},
\hspace{0.4cm}
\rho^{\alpha}_{\mu\tau}=\begin{pmatrix}
\rho_{55}^{\alpha}&0&0&0\\
0&\rho_{22}^{\alpha}&\rho_{23}^{\alpha}&0\\
0&\rho_{32}^{\alpha}&\rho_{33}^{\alpha}&0\\
0&0&0&0
\end{pmatrix}.
\label{17}
\end{equation}

By tracing again we can obtain the reduced density matrices of the single-partite subsystems:
\begin{equation}
\rho^{\alpha}_{e}=\begin{pmatrix}
\rho_{22}^{\alpha}+\rho_{33}^{\alpha}&0\\
0&\rho_{55}^{\alpha}\\
\end{pmatrix},
\hspace{0.4cm}
\rho^{\alpha}_{\mu}=\begin{pmatrix}
\rho_{22}^{\alpha}+\rho_{55}^{\alpha}&0\\
0&\rho_{33}^{\alpha}\\
\end{pmatrix},
\hspace{0.4cm}
\rho^{\alpha}_{\tau}=\begin{pmatrix}
\rho_{55}^{\alpha}+\rho_{33}^{\alpha}&0\\
0&\rho_{22}^{\alpha}\\
\end{pmatrix}.
\label{18}
\end{equation}

By following the above prescription,  it is simple to evaluate the CCR terms of  Eq.(\ref{9}):
\begin{eqnarray}
P_{\rm{hs}}(\rho^{\alpha}_{e})&=&(|a_{\alpha\mu}(t)|^{2}+|a_{\alpha\tau}(t)|^{2})^{2}+|a_{\alpha e}(t)|^{2}-\frac{1}{2},\\
C_{\rm{hs}}(\rho^{\alpha}_{e})&=&0,\\
C_{\rm{hs}}^{nl}(\rho^{\alpha}_{e|\mu\tau})&=&1-|a_{\alpha e}(t)|^{2}-(|a_{\alpha\mu}(t)|^{2}+|a_{\alpha\tau}(t)|^{2})^{2}.
\end{eqnarray}
By summing up all these terms we verify that Eq.(\ref{9}) is satisfied.

For a state such as  in Eq.(\ref{13}),  $C_{\rm{hs}}^{nl}(\rho^{\alpha}_{e|\mu\tau})=C_{hs}(\rho^{\alpha}_{e\mu})+C_{hs}(\rho^{\alpha}_{e\tau})$, i.e. the non-local coherence that the subsystem $e$ shares with $\mu\tau$ is equal to the sum of the bipartite correlations that $e$ shares with $\mu$ and $\tau$ separately. So, Eq.(\ref{9}) can be written as:
\begin{equation}
P_{\rm{hs}}(\rho^{\alpha}_{e})+C_{hs}(\rho^{\alpha}_{e\mu})+C_{hs}(\rho^{\alpha}_{e\tau})=\frac{1}{2},
\label{22}
\end{equation}
with $C_{hs}(\rho^{\alpha}_{e\mu})\!=\!(a_{\alpha e}(t)a_{\alpha \mu}(t)^{*})^{2}\!+\!(a_{\alpha \mu}(t)a_{\alpha e}^{*}(t))^{2}$ and $C_{hs}(\rho^{\alpha}_{e\tau})\!=\!(a_{\alpha e}(t)a_{\alpha \tau}^{*}(t))^{2}\!+\!(a_{\alpha\tau}(t)a_{\alpha e}^{*}(t))^{2}$.

Let us now evaluate the terms of Eq.(\ref{11}) for subsystem $e\mu$. By evaluating the eigenvalues of the reduced density matrices in Eq.(\ref{17}) we obtain:
\begin{eqnarray}
S_{vn}(\rho^{\alpha}_{e\mu})&=&-(P_{\alpha e}+P_{\alpha\mu})\log_{2}(P_{\alpha e}+P_{\alpha \mu})-P_{\alpha \tau}\log_{2}P_{\alpha\tau},\\
P_{vn}(\rho^{\alpha}_{e\mu})&=&2+P_{\alpha e}\log_{2}P_{\alpha e}+P_{\alpha\mu}\log_{2}P_{\alpha\mu}+P_{\alpha\tau}\log_{2}P_{\alpha\tau},\\
C_{re}(\rho^{\alpha}_{e\mu})&=&-P_{\alpha e}\log_{2}P_{\alpha e}-P_{\alpha\mu}\log_{2}P_{\alpha\mu}+(P_{\alpha e}+P_{\alpha\mu})\log_{2}(P_{\alpha e}+P_{\alpha\mu}).
\end{eqnarray}

Similar expressions have been obtained for subsystems $e\tau$ and $\mu\tau$.

\begin{figure}
\begin{minipage}{18pc}
\includegraphics[width=18pc]{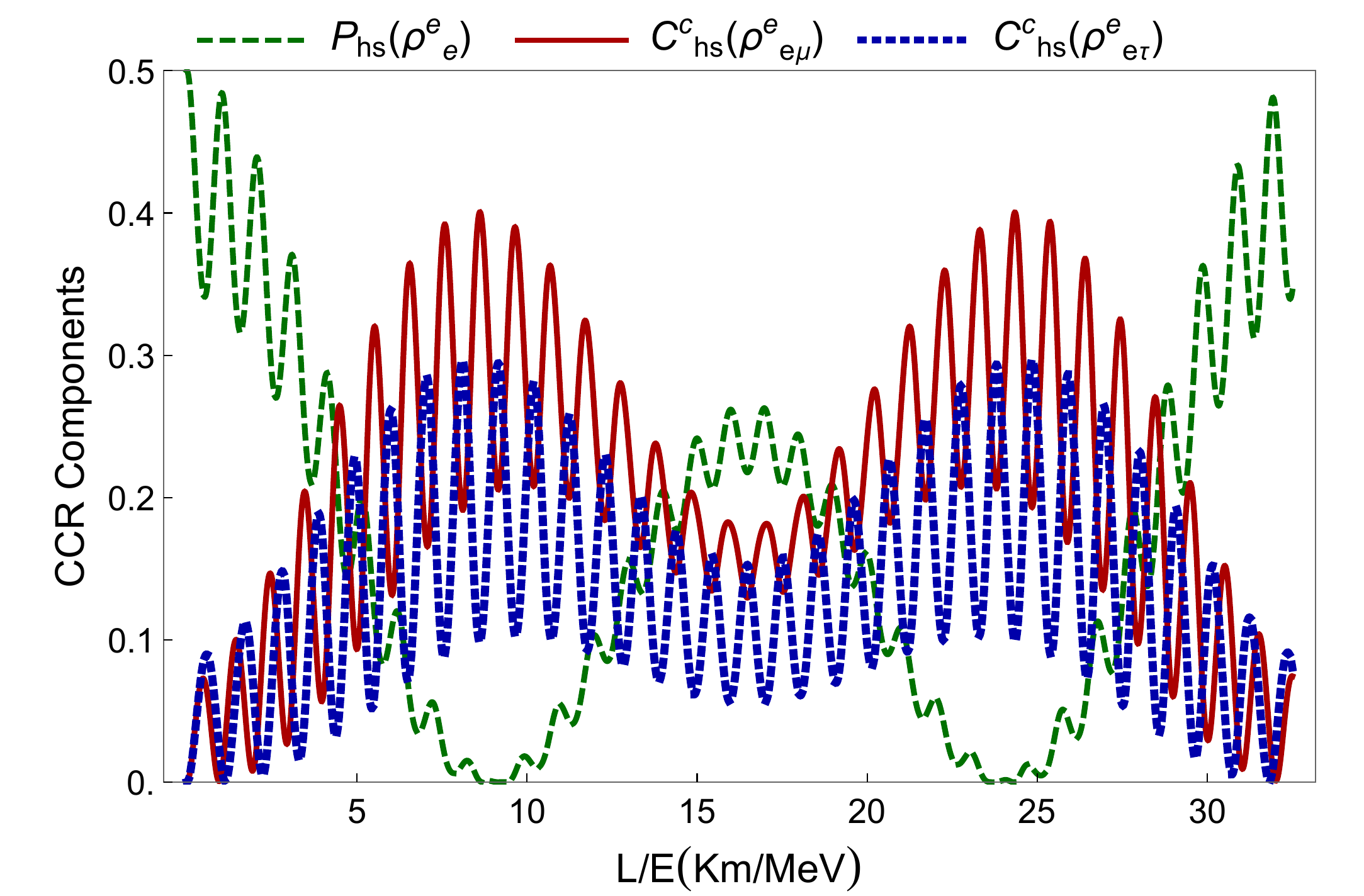}
\caption{\label{figure1}CCR terms, for an initial electronic neutrino, of Eq.(\ref{22}) as function of $L/E$.}
\end{minipage}\hspace{2pc}%
\begin{minipage}{18pc}
\includegraphics[width=18pc]{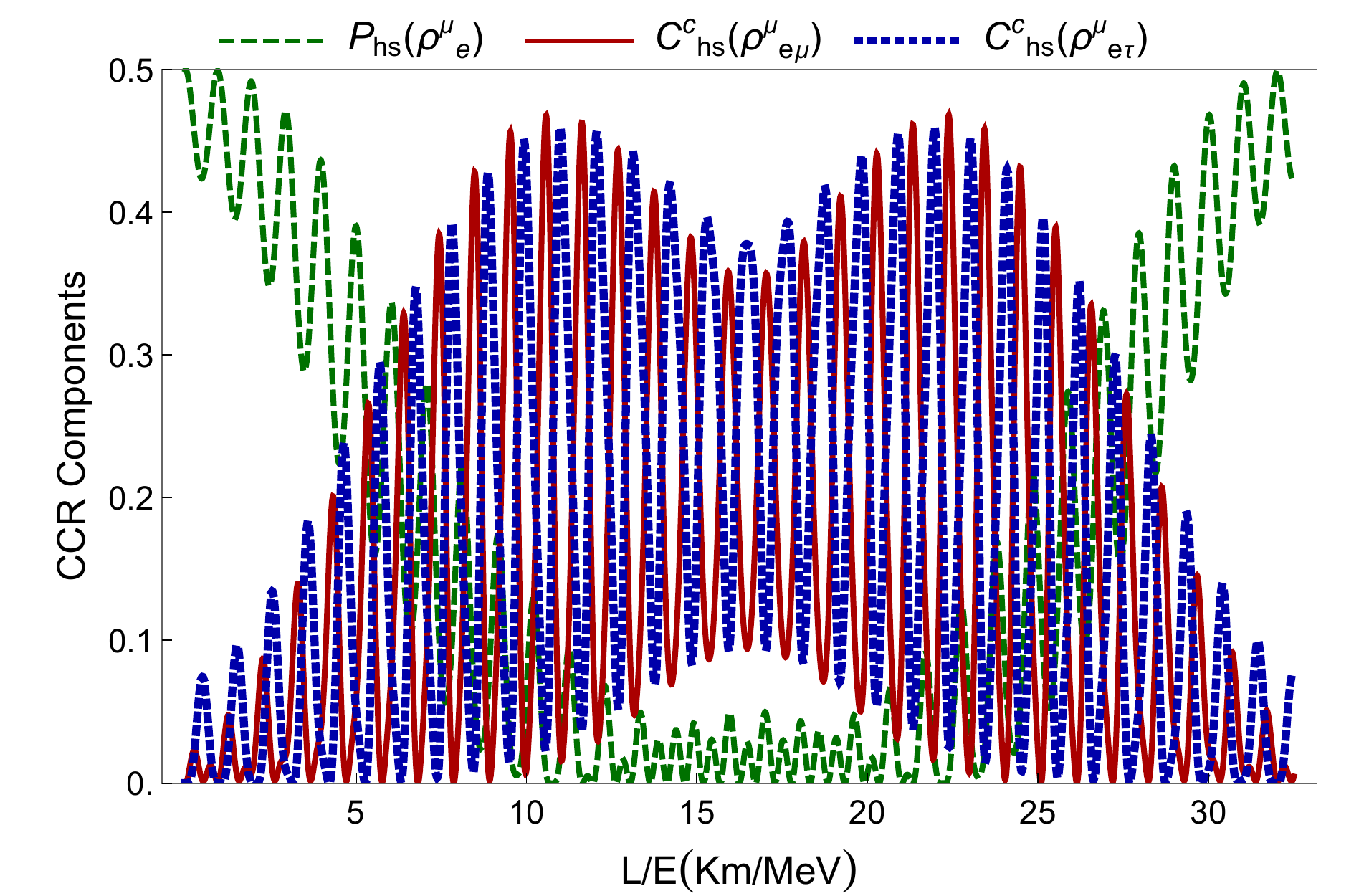}
\caption{\label{figurem1}CCR terms, for an initial muonic neutrino, of Eq.(\ref{22}) as function of $L/E$.}
\end{minipage}
\end{figure}
\newpage

\begin{figure}[h]
\begin{minipage}{18pc}
\subfloat[][\emph{$e\mu$ subsystem }]{\includegraphics[width=18pc]{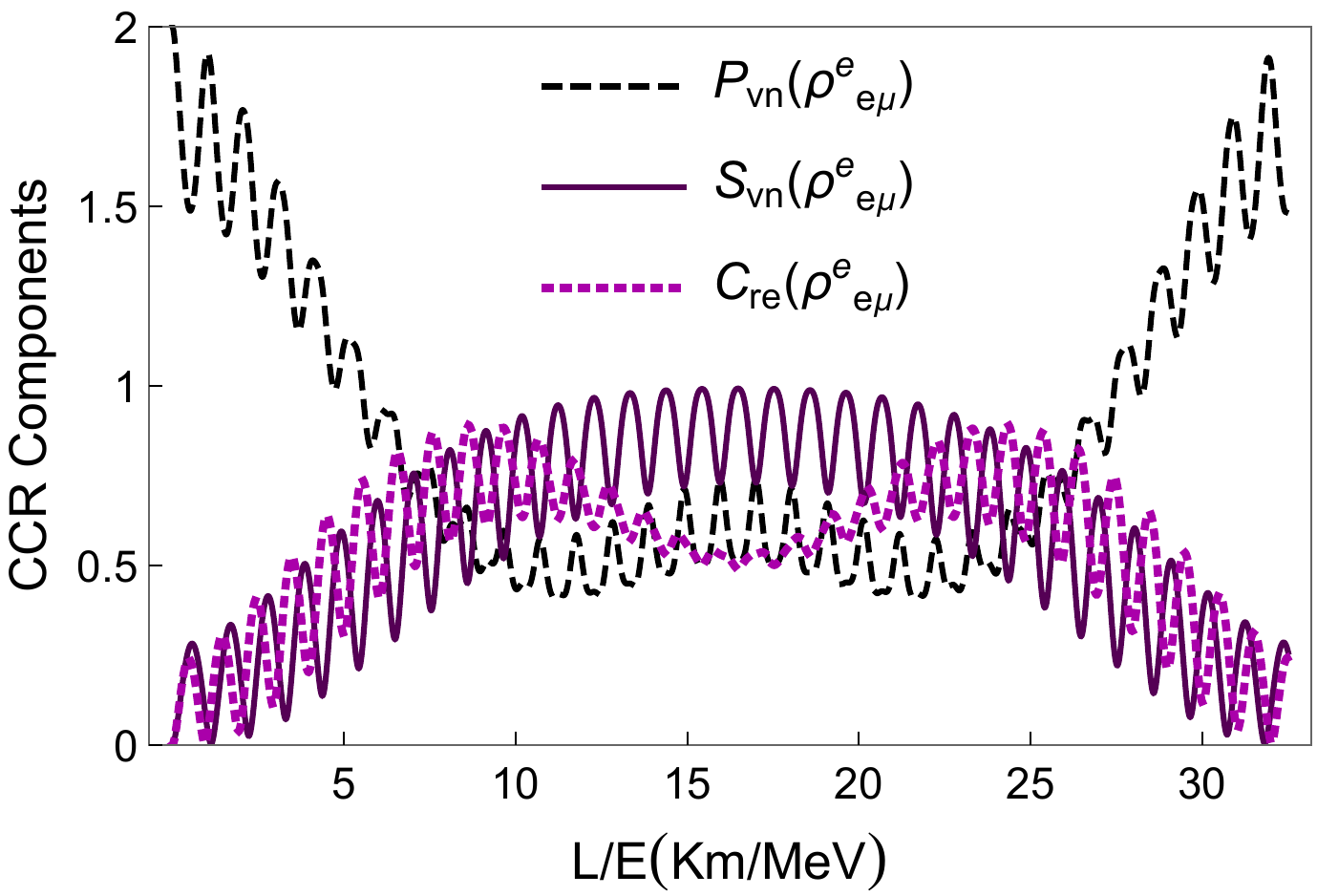}}\\
\subfloat[][\emph{$e\tau$ subsystem }]{\includegraphics[width=18pc]{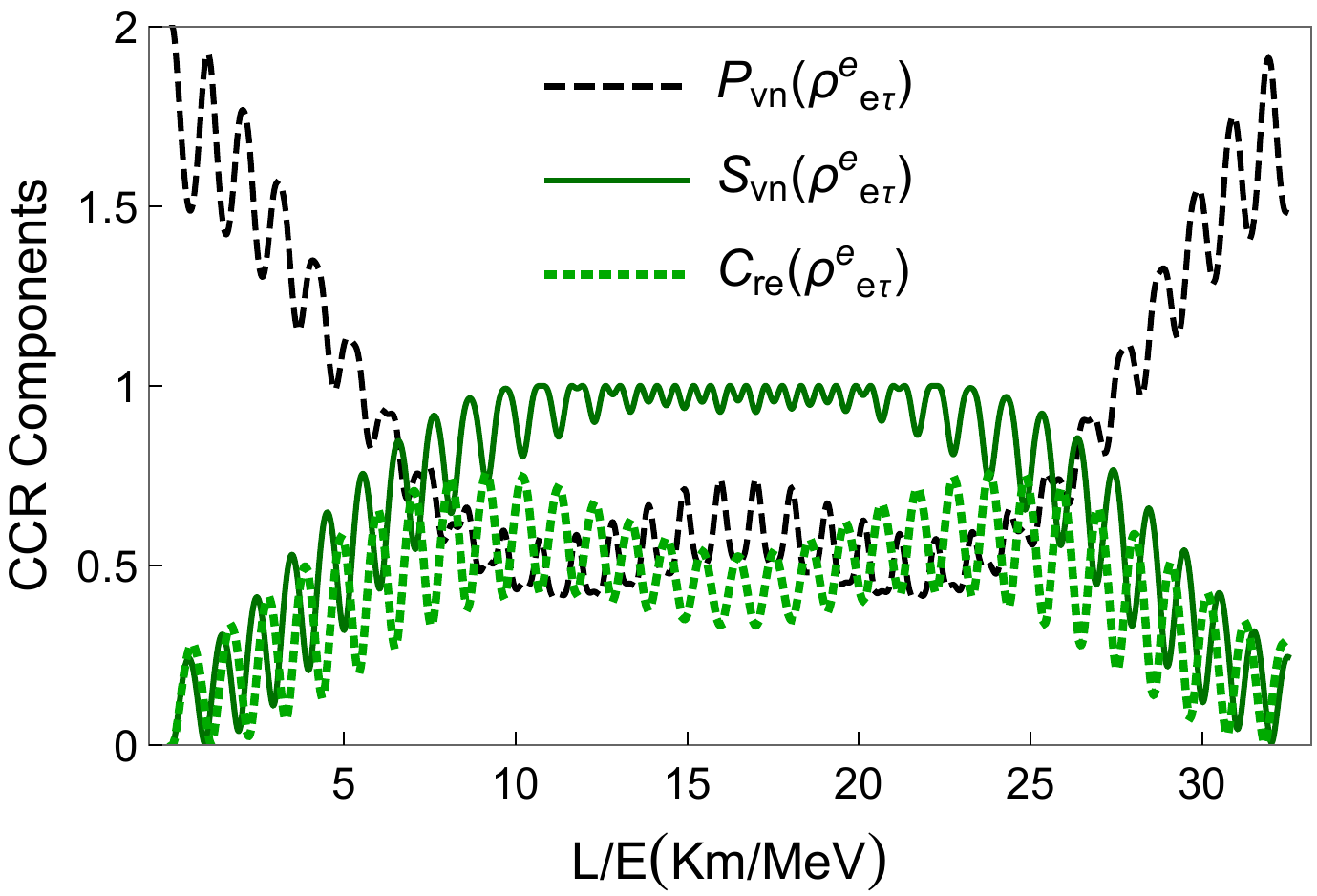}}\\
\subfloat[][\emph{$\mu\tau$ subsystem }]{\includegraphics[width=18pc]{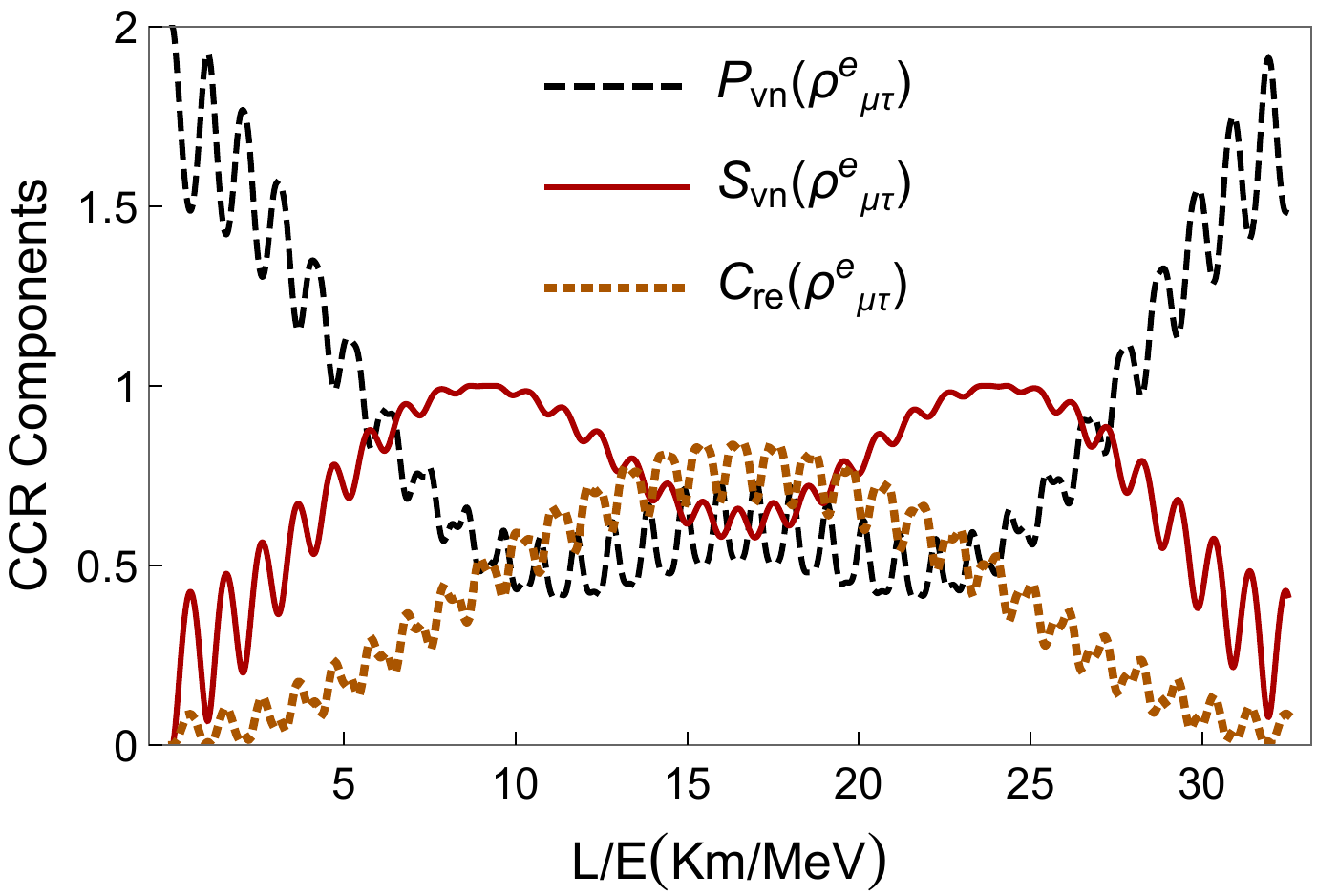}}
\caption{\label{figure2}CCR terms for bipartite subsystems $e\mu$, $e\tau$ and  $\mu\tau$ as function of $L/E$ in the case of an initial electronic neutrino.}
\end{minipage}\hspace{2pc}%
\begin{minipage}{18pc}
\subfloat[][\emph{$e\mu$ subsystem }]{\includegraphics[width=18pc]{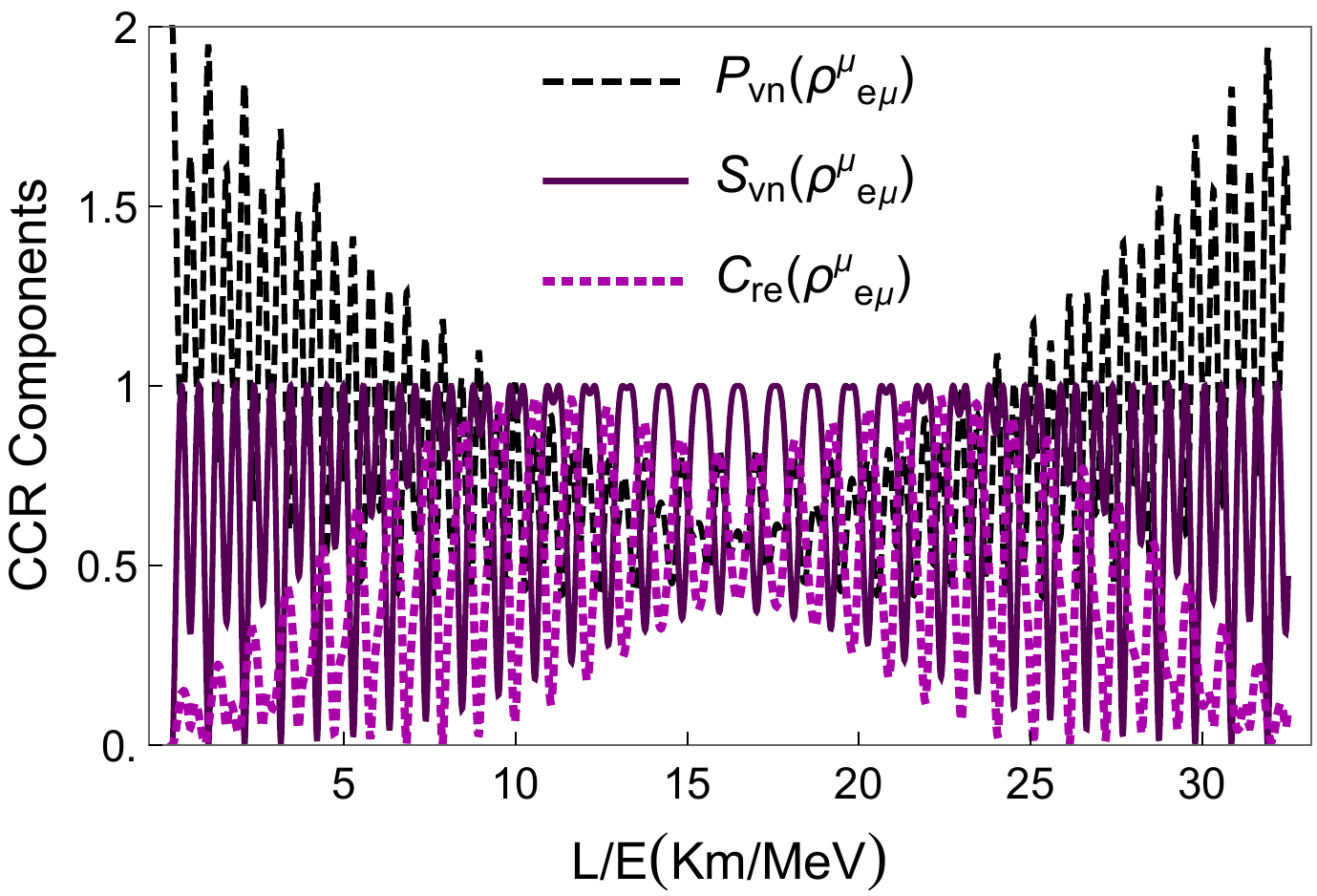}}\\
\subfloat[][\emph{$e\tau$ subsystem }]{\includegraphics[width=18pc]{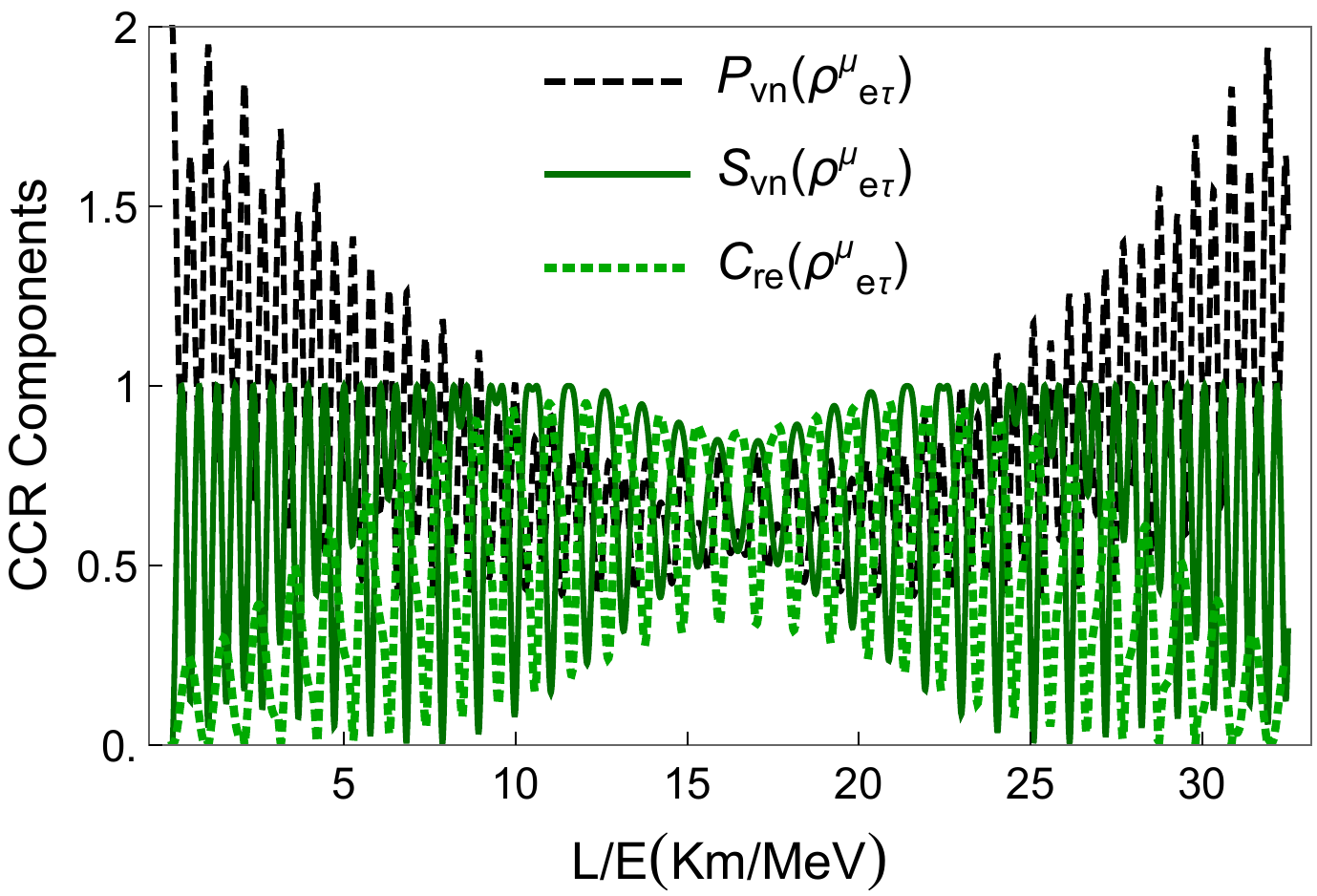}}\\
\subfloat[][\emph{$\mu\tau$ subsystem }]{\includegraphics[width=18pc]{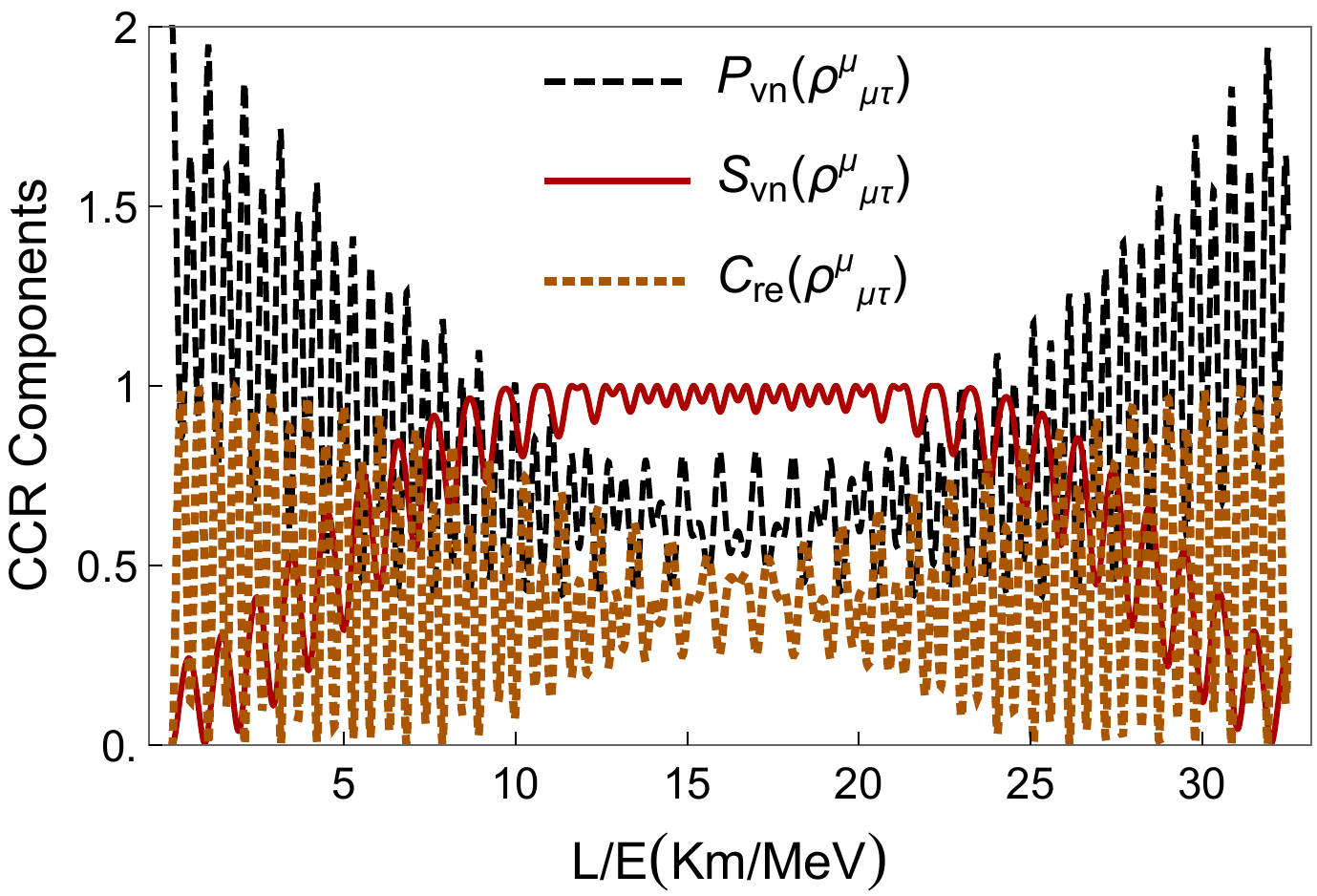}}
\caption{\label{figurem2}CCR terms for bipartite subsystems $e\mu$, $e\tau$ and  $\mu\tau$ as function of $L/E$ in the case of an initial muonic neutrino.}

\end{minipage} 
\end{figure}

\begin{figure}[h]
{\includegraphics[width=18pc]{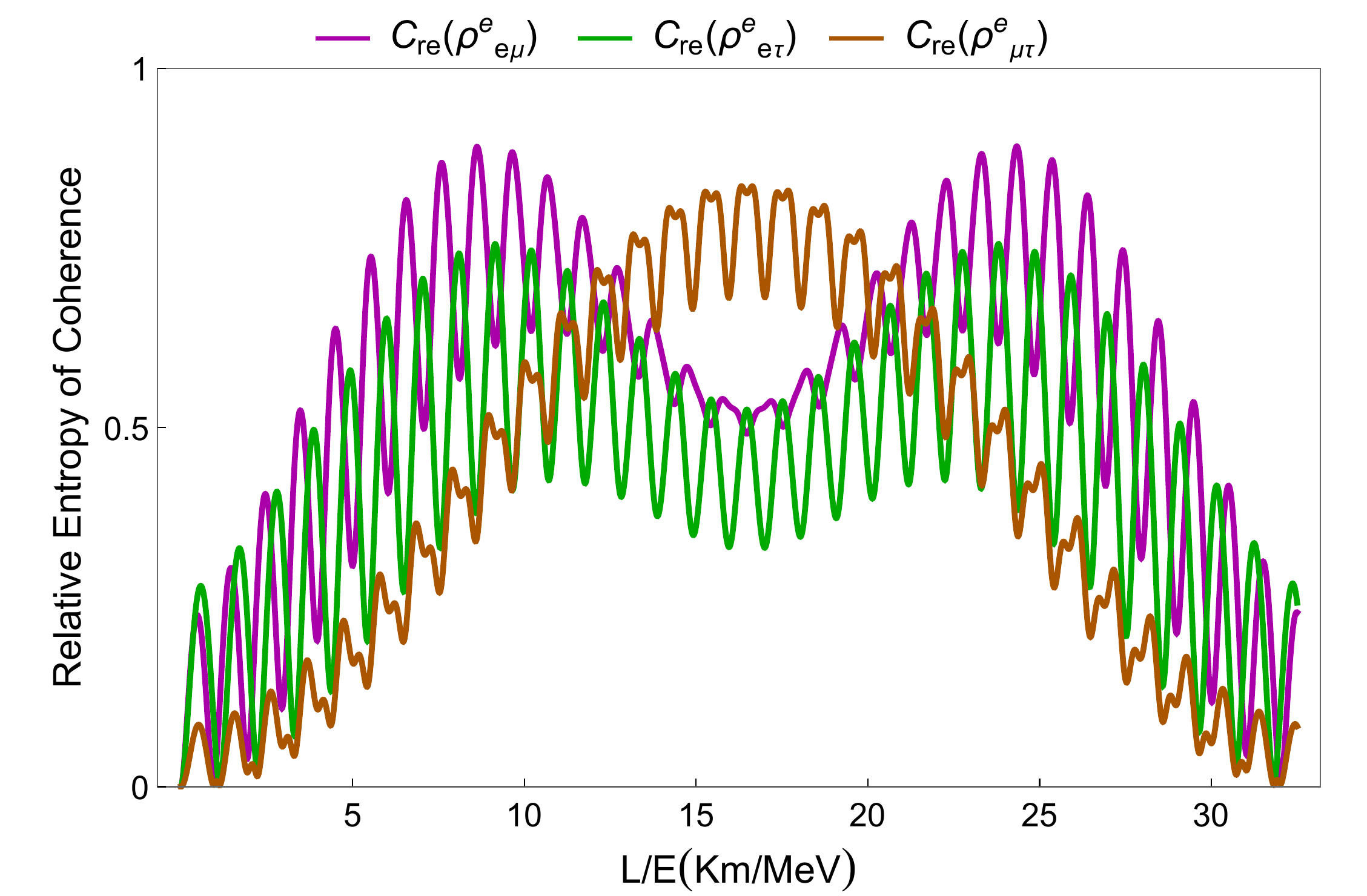}}\quad
{\includegraphics[width=18pc]{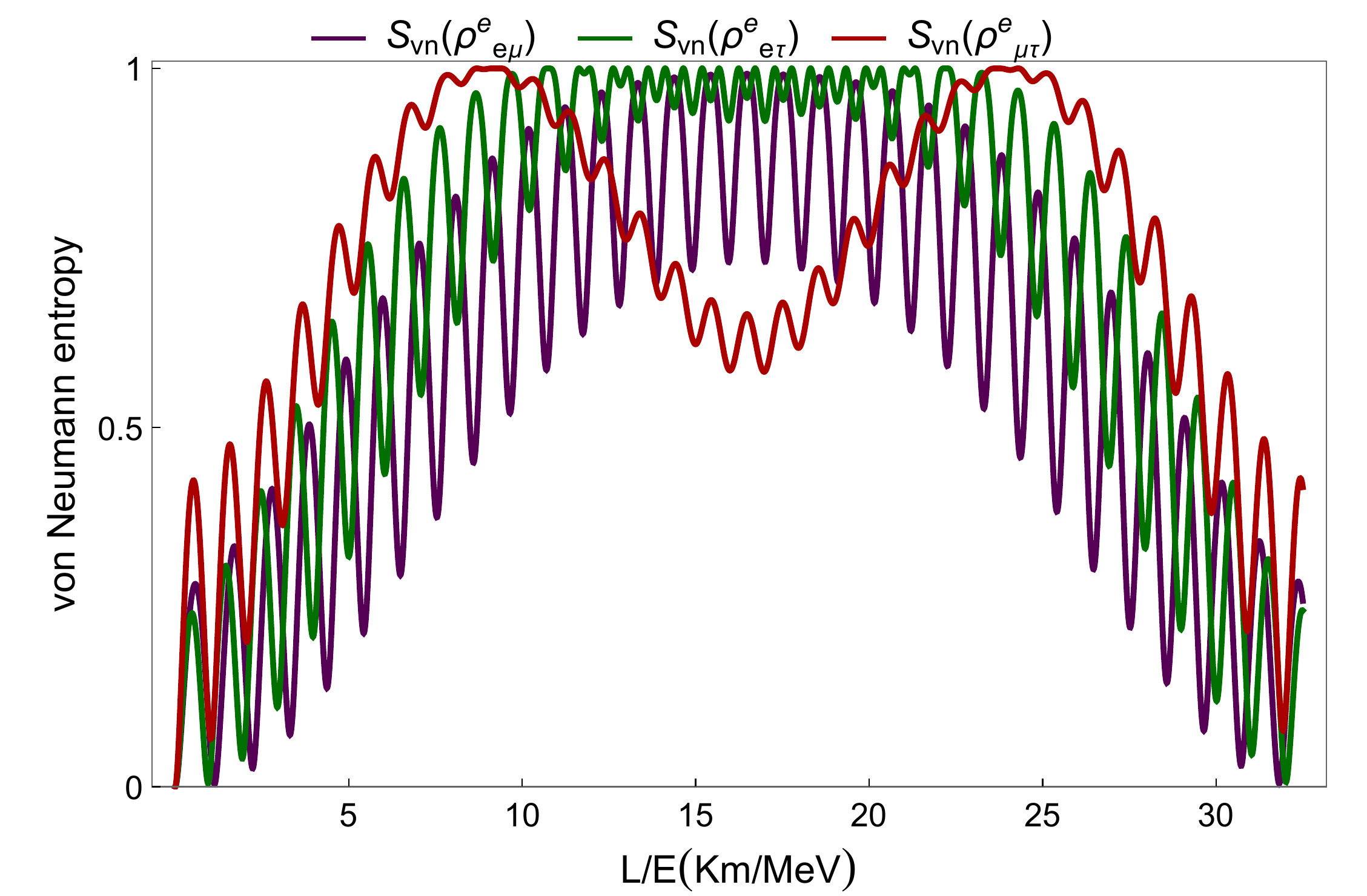}}
\caption{\label{figure3}Comparison among $C_{re}(\rho^{e}_{e\mu})$, $C_{re}(\rho^{e}_{e\tau})$ and $C_{re}(\rho^{e}_{\mu\tau})$ (left panel) and  $S_{vn}(\rho^{e}_{e\mu})$, $S_{vn}(\rho^{e}_{e\tau})$ and $S_{vn}(\rho^{e}_{\mu\tau})$ (right panel), for an initial electronic neutrino. }
\end{figure}

\begin{figure}[h]
{\includegraphics[width=18pc]{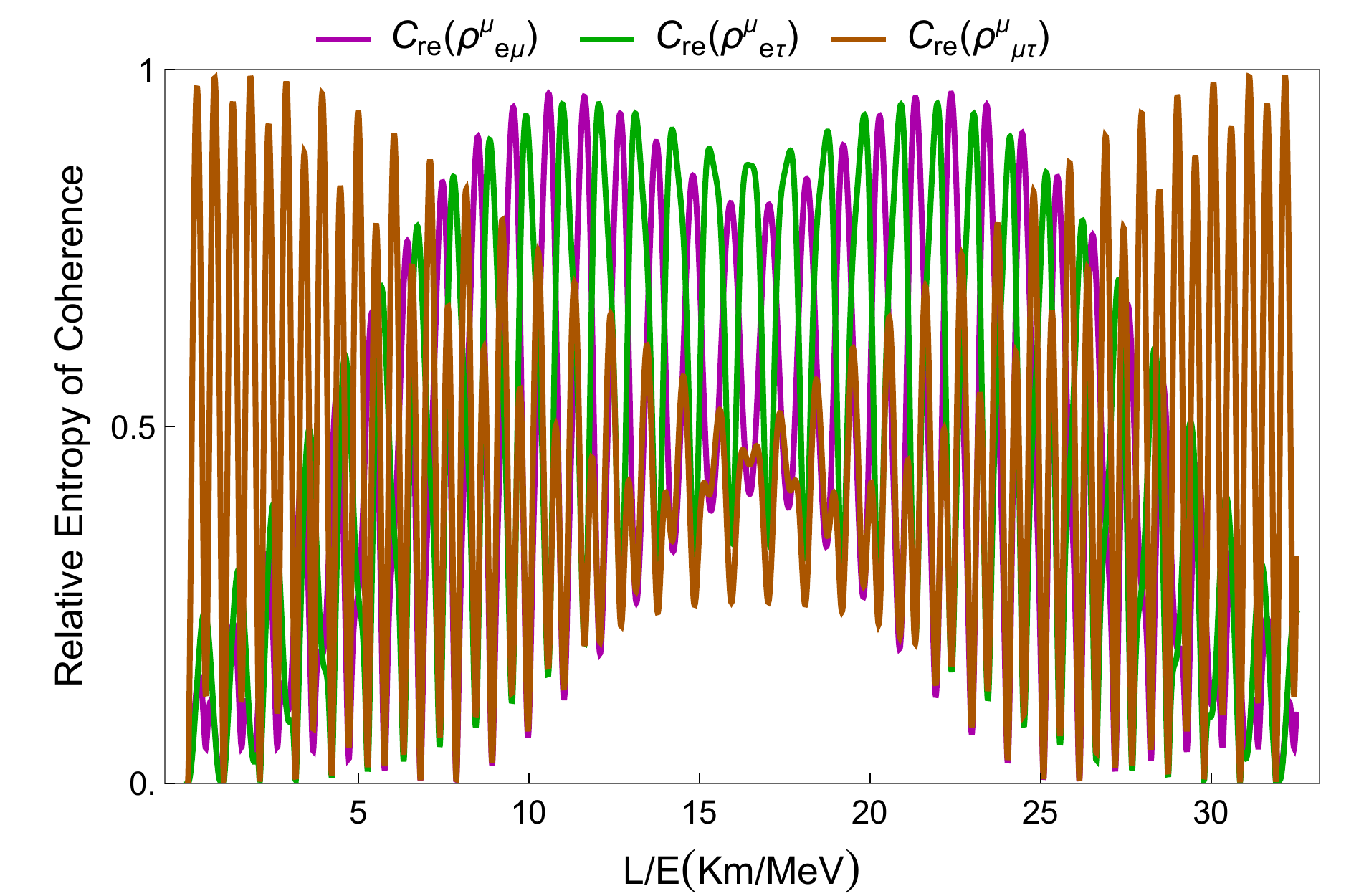}}\quad
{\includegraphics[width=18pc]{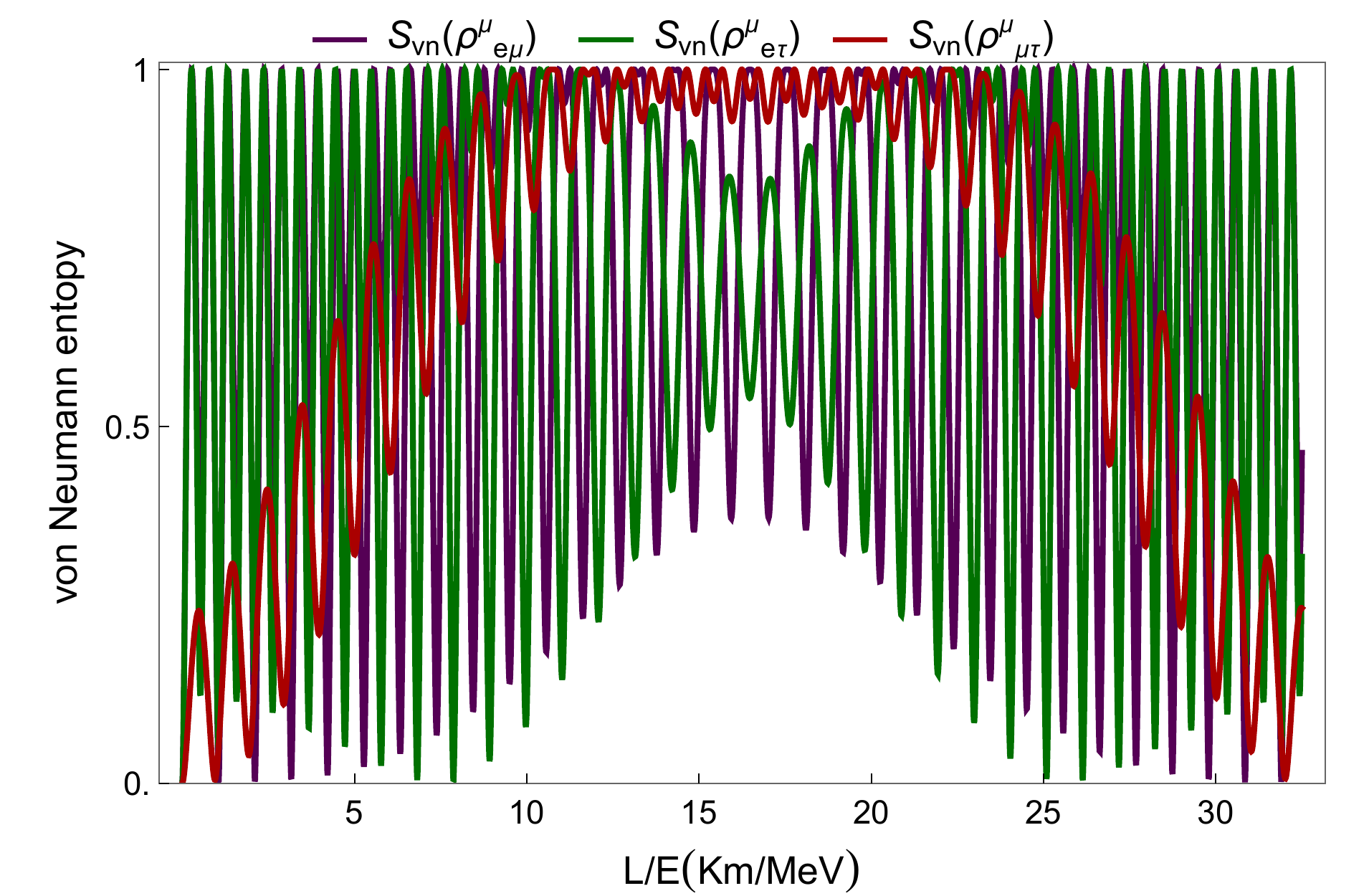}}
\caption{\label{figurem3}Comparison among $C_{re}(\rho^{\mu}_{e\mu})$, $C_{re}(\rho^{\mu}_{e\tau})$ and $C_{re}(\rho^{\mu}_{\mu\tau})$ (left panel) and  $S_{vn}(\rho^{\mu}_{e\mu})$, $S_{vn}(\rho^{\mu}_{e\tau})$ and $S_{vn}(\rho^{\mu}_{\mu\tau})$ (right panel), for an initial muonic neutrino.}
\label{figurem3}
\end{figure}

\newpage

\subsubsection{Behavior of the CCR components for electron and muon neutrinos}

\textit{A. Electronic case}\\

At first we plot the  terms of Eqs.(\ref{22},\ref{11}) as function of $L/E$ for $\alpha=e$. Therefore, we will comment here figures \ref{figure1}, \ref{figure2}, \ref{figure3}.

Referring to Eq.(\ref{22}) in Fig.(\ref{figure1}) we notice that the bipartite correlations shared between $e$ and $\mu$ are larger than those shared between $e$ and $\tau$. Therefore, the coherence term $C_{hs}(\rho^{e }_{e\mu})$ gives a more prominent contribution in completing the complementary relation with respect to $C_{hs}(\rho^{e }_{e\tau})$. Similar considerations are valid for the case of an initial state of flavor $\mu$ or $\tau$. 

Referring to Eq.(\ref{11})in Fig.\ref{figure2} we show that, interestingly, the von Neumann entropy exhibits a plateau (see Fig.\ref{figure2}(a)), which is in correspondence of its maximum value for the $e\tau$ subsystem. Such a behavior persists for a relatively large range of L/E.

In Fig.\ref{figure3}  we compare the three bipartite local coherences $C_{re}(\rho^{e}_{e\mu})$, $C_{re}(\rho^{e}_{e\tau})$ and $C_{re}(\rho^{e}_{\mu\tau})$ and the von Neumann entropies $S_{vn}(\rho^{e}_{e\mu})$, $S_{vn}(\rho^{e}_{e\tau})$ and $S_{vn}(\rho^{e}_{\mu\tau})$, above reported in Fig.(\ref{figure2}).The dynamical behavior of the different correlations depend strongly on the bipartition considered, indicating the role of true tripartite correlations.


\medskip

\noindent\textit{B. Muonic case}

Here we plot the terms of Eqs. (\ref{22}, \ref{11})
as function of $L/E$ for $\alpha=\mu$, and we refer to figures \ref{figurem1},\ref{figurem2},\ref{figurem3}.

In  Fig.\ref{figurem1} we show the terms of Eq.(\ref{22}) and we observe that, differently to the electron case, it is difficult to recognize a dominant contribution of one between $C_{hs}(\rho^{\mu}_{e\mu})$ and $C_{hs}(\rho^{\mu}_{e\tau})$,
while we recognize a sort of anti-correlation between them.

In Fig.\ref{figurem2} we plot the CCR terms of Eq.(\ref{11}) for subsystems $e\mu$,  $e\tau$ and $\mu\tau$. Similar to the previous case, the von Neumann entropy exhibits a plateau which is in the $\mu\tau$ subsystem (Fig.\ref{figurem2}(c)).

We also show a comparison of the three bipartite local coherences and the von Neumann entropies of each bipartition in Fig.\ref{figurem3}. Again, it is striking how the behavior of the correlations depends on the specific bipartition.



\section{CCR for mixed systems}

The CCR used so far are only valid for pure states. We now revise the extension of CCR to mixed states, as was first derived in \cite{Basso21}. This will allow us to analyze quantum correlations in neutrino oscillations when the wave-packet treatment is incorporated. In this case, a flavor state is a superposition of wave-packets that propagate with different group velocities, so inducing decoherence and increasing the mixedness of the state as it propagates.


\subsection{CCR for bipartite states}
For mixed states, CCR have to be modified to correctly quantify the complementarity behaviour of subsystems. In fact, in this case the term $S_{vn}(\rho_{A})$ in Eq.(\ref{6})  cannot be considered as a measure of entanglement, but it is just a measure of mixedness of A.
The correct form of CCR to consider for bipartite mixed states is given by \cite{Basso21}:
\begin{equation}
I_{A:B}(\rho_{AB})+S_{A|B}(\rho_{AB})+P_{vn}(\rho_{A})+C_{re}(\rho_{A})=
\log_{2}(d_{A}),
\label{26}
\end{equation}
where $I_{A:B}(\rho_{AB})$ is the mutual information of A and B and $S_{A|B}(\rho_{AB})=S_{vn}(\rho_{AB})-S_{vn}(\rho_{B})$ quantifies the ignorance about the whole system we have by looking only to subsystem A.
\subsection{CCR for tripartite states}
Similar considerations are valid for the case of tripartite mixed state. If in Eq.(\ref{26}) we make $A\rightarrow (AB)$, CCR for the subsystem AB takes the form:
\begin{equation}P_{vn}(\rho_{AB})+C_{re}(\rho_{AB})+S_{AB|C}(\rho_{ABC})+I_{AB:C}(\rho_{ABC})=\log_{2}(d_{A}d_{B}),
\label{27}
\end{equation}
where $I_{AB:C}(\rho_{ABC})=S_{{\rm{vn}}}(\rho_{AB})+S_{{\rm{vn}}}(\rho_{C})-S_{{\rm{vn}}}(\rho_{ABC})$ and $S_{AB|C}(\rho_{ABC})=S_{{\rm{vn}}}(\rho_{ABC})-S_{{\rm{vn}}}(\rho_{C})$.

The state for the subsystem C, on the other hand, satisfy the CCR:
\begin{equation}
P_{vn}(\rho_{C})+C_{re}(\rho_{C})+S_{C|AB}(\rho_{ABC})
+I_{C:AB}(\rho_{ABC})=\log_{2}(d_{C}).
\label{28}
\end{equation}

In \cite{CCR2} it has been shown that for the case of a bipartite mixed state, the sum of the two non-local terms of the CCR for mixed state results to be equal to the quantum discord. This will remain valid for a mixed tripartite state.

\subsection{CCR for tri-partite mixed neutrino states}
Let us suppose to have,  at $t=0$, a neutrino (mixed) state with flavor $\alpha=e,\mu$. The density matrix associated to this state (see Eq.(\ref{n6}) of the Appendix \ref{A}), is:
\begin{equation}
\rho_{e\mu\tau}^{\alpha}=
\begin{pmatrix}
0&0&0&0&0&0&0&0\\
0&F_{\tau\tau}^{\alpha}&F_{\tau\mu}^{\alpha}&0&F_{\tau e}^{\alpha}&0&0&0\\
0&F_{\mu\tau}^{\alpha}&F_{\mu\mu}^{\alpha}&0&F_{\mu e}^{\alpha}&0&0&0\\
0&0&0&0&0&0&0&0\\
0&F_{e \tau}^{\alpha}&F_{e\mu}^{\alpha}&0&F_{ee}^{\alpha}&0&0&0\\
0&0&0&0&0&0&0&0\\
0&0&0&0&0&0&0&0\\
0&0&0&0&0&0&0&0
\end{pmatrix}
\label{29}
\end{equation}
where $F^{\alpha}_{\beta\gamma}$ are given in  Appendix \ref{A}. By evaluating the reduced density matrices for the bipartite and single-partite subsystems, it is possible to compute the terms of Eq.(\ref{27}).

The terms of the CCR for tri-partite mixed states can be evaluated as functions of the different $F^{\alpha}_{\beta\gamma}$ appearing in Eq.~\eqref{29} and their explict expressions
 for the subsystem $e \mu$ are reported in Appendix \ref{B}.

 Before prooceeding we specify that, in what follows, we will use the following oscillation parameters, which appear in recent neutrino experiments\cite{Acero}--\cite{Abe}:
\begin{equation}
\begin{split}
\Delta m_{21}^{2}=7.50\times10^{-5}eV^{2},\\
\Delta m_{31}^{2}=2.46\times10^{-3}eV^{2},\\
\Delta m_{32}^{2}=2.38\times10^{-3}eV^{2},\\
\theta_{12}=33.48^{\circ},\theta_{23}=42.3^{\circ},\theta_{13}=8.50^{\circ}.
\end{split}
\label{n5}
\end{equation}
For simplicity, here we consider $\delta_{CP}=0$.

\subsubsection{Results for electron neutrino oscillations}

\begin{figure}[t]
\begin{minipage}{17cm}
\subfloat[]{
\includegraphics[width=5.5cm]{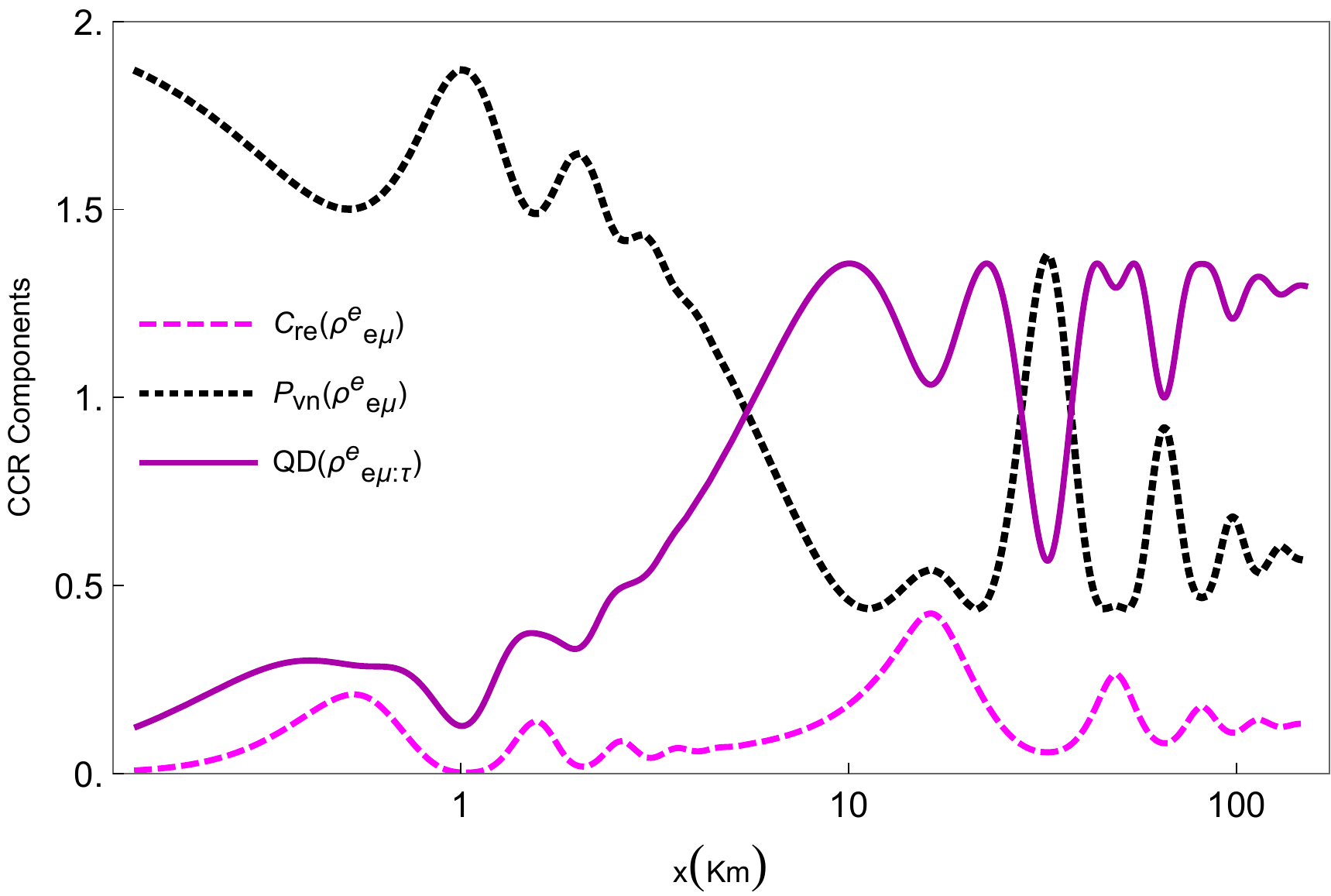}
}
\subfloat[]{
\includegraphics[width=5.5cm]{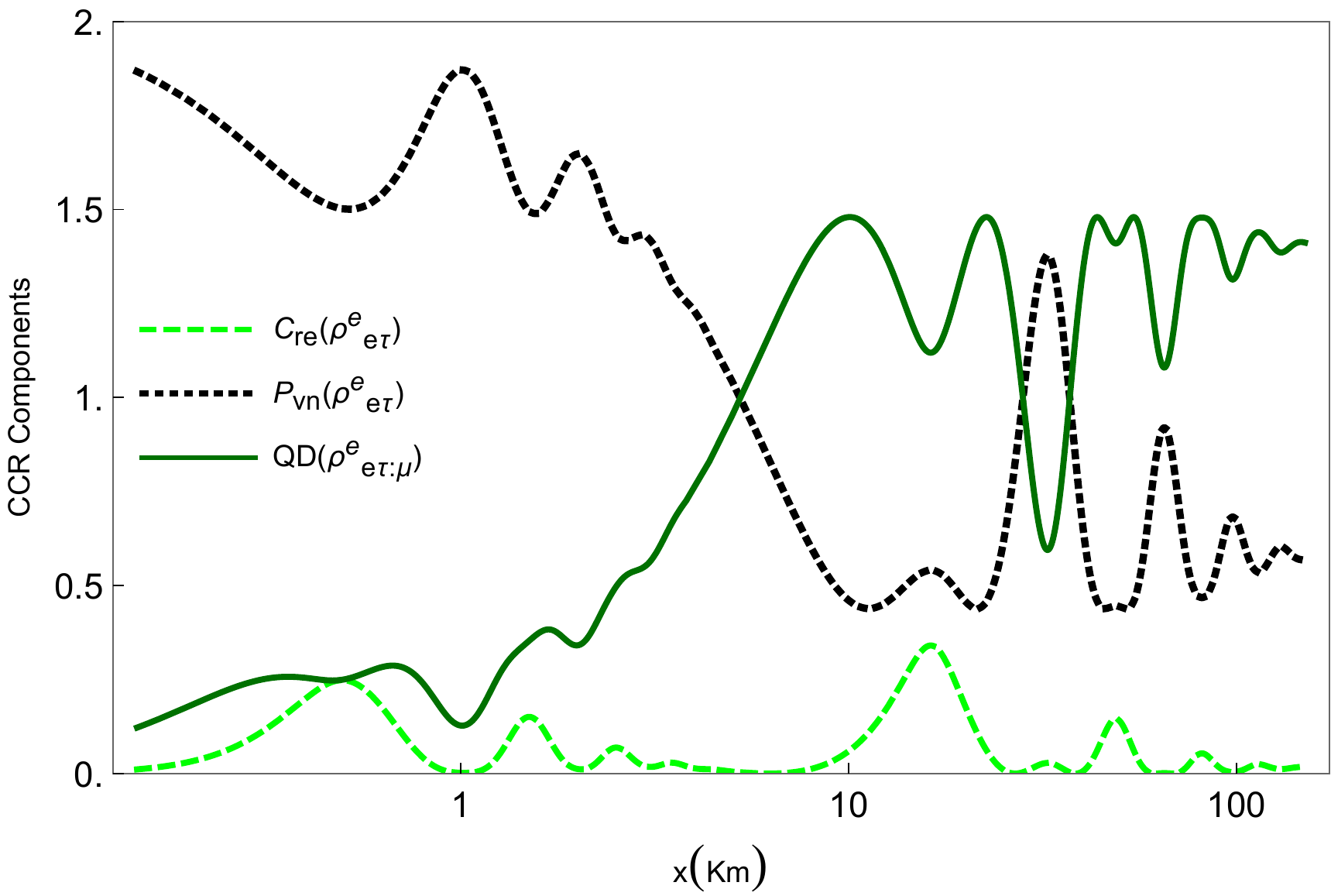}
}
\subfloat[]{
\includegraphics[width=5.5cm]{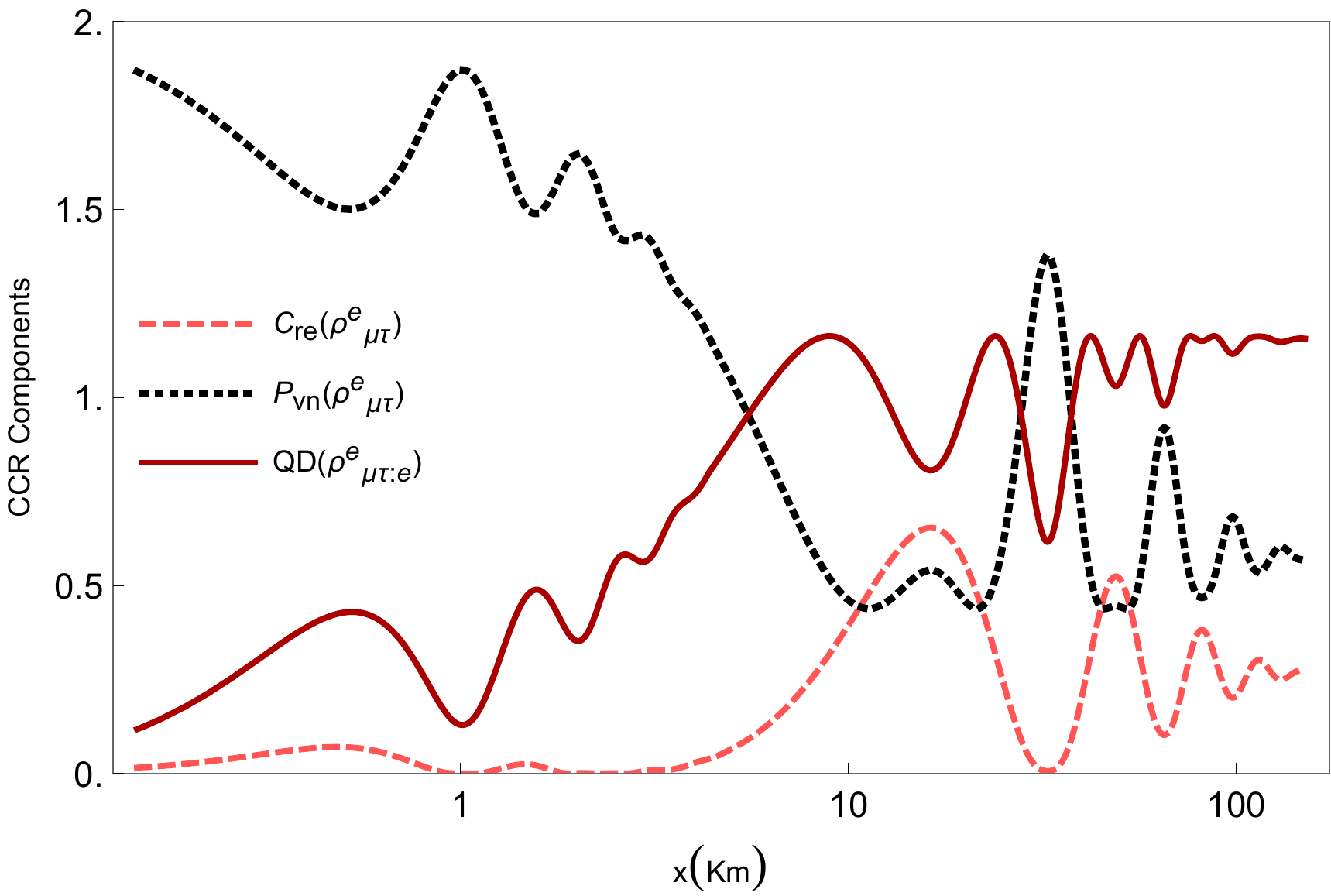}
}
\caption{CCR terms for two flavor subsystems $e\mu$ (a), $e\tau$ (b) and  $\mu\tau$ (c) as function of $x$, in the case of an initial electron neutrino state.}
\label{figure7}
\end{minipage}
\end{figure}

In Fig.\ref{figure7} are shown the CCR terms of Eq.(\ref{27}), for a neutrino system, as function of the distance, for the three possible two flavor subsystems $e\mu$, $e\tau$ and $\mu\tau$. As above remarked, one can see that the behavior of these terms is different depending on the bipartite subsystem considered. 

At great distances, the dominant contribution to the correlations is given by the Quantum Discord (given by the sum of the last two terms of left hand-side of   Eq.(\ref{27}). A small contribution is provided from the internal coherence of the two-flavor subsystems, and represents  a ``local''\footnote{We stress that here ``local'' is referred to flavor, i.e. restricted to only two flavors, in contrast to the ``global'' three flavor state} coherence that is not present for a single-partite state, as shown in \cite{CCR2}.

Apart from the predictability, which behaves exactly the same for all bipartitions, the ``local'' coherences, associated to $e\mu$, $e\tau$ and $\mu\tau$ subsystems, change depending on the subsystem. In particular, $C_{re}(\rho^{e}_{e\tau})$, which is greater than $C_{re}(\rho^{e}_{\mu\tau})$ and slightly larger than $C_{re}(\rho^{e}_{e\mu})$ at small distances, becomes close to zero at large $x$, where the maximum coherence is given by $C_{re}(\rho^{e}_{\mu\tau})$.

\smallskip

\subsubsection{Result for muon neutrino oscillations}

\begin{figure}[t]
\begin{minipage}{17cm}
\subfloat[]{
\includegraphics[width=5.5cm]{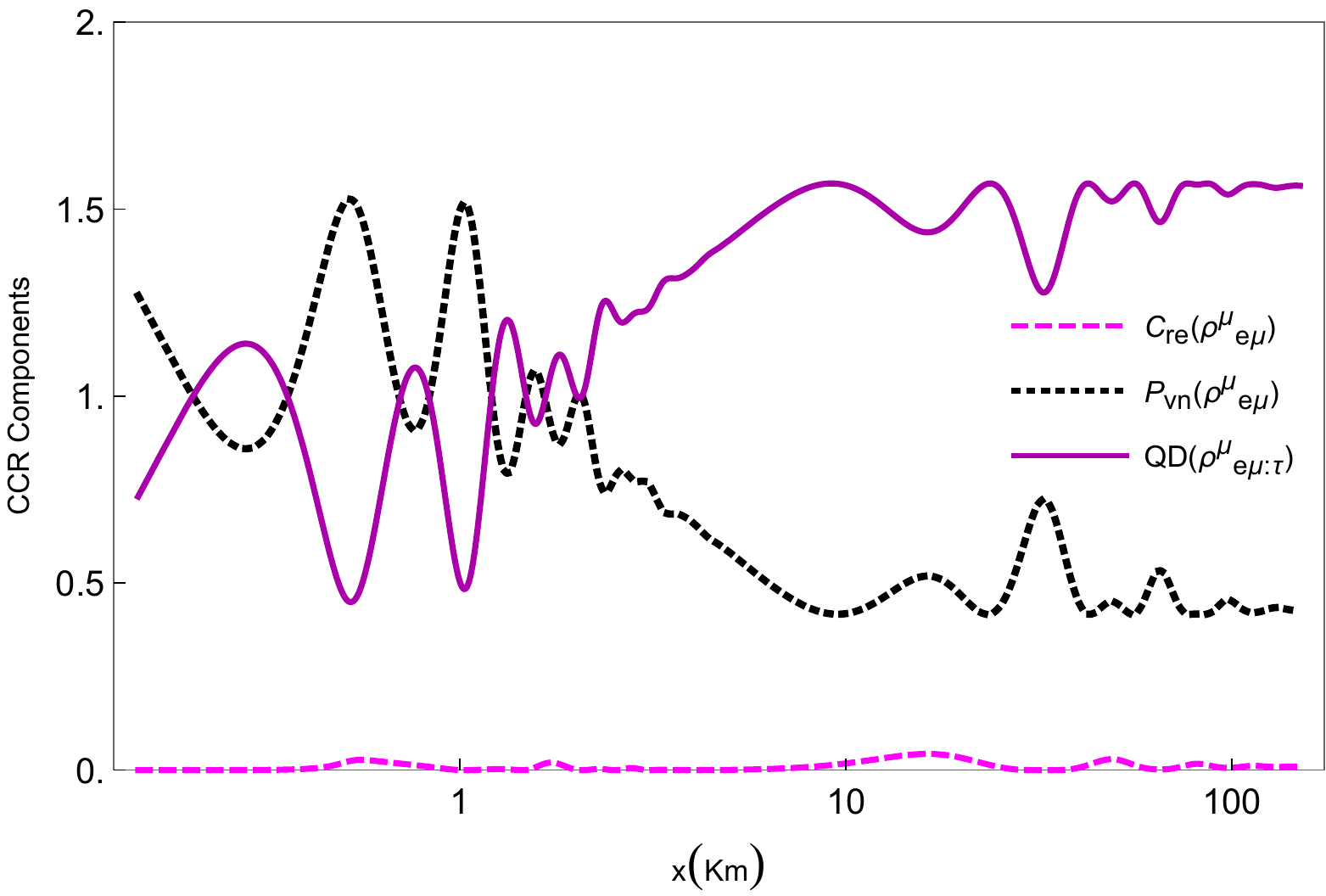}
}
\subfloat[]{
\includegraphics[width=5.5cm]{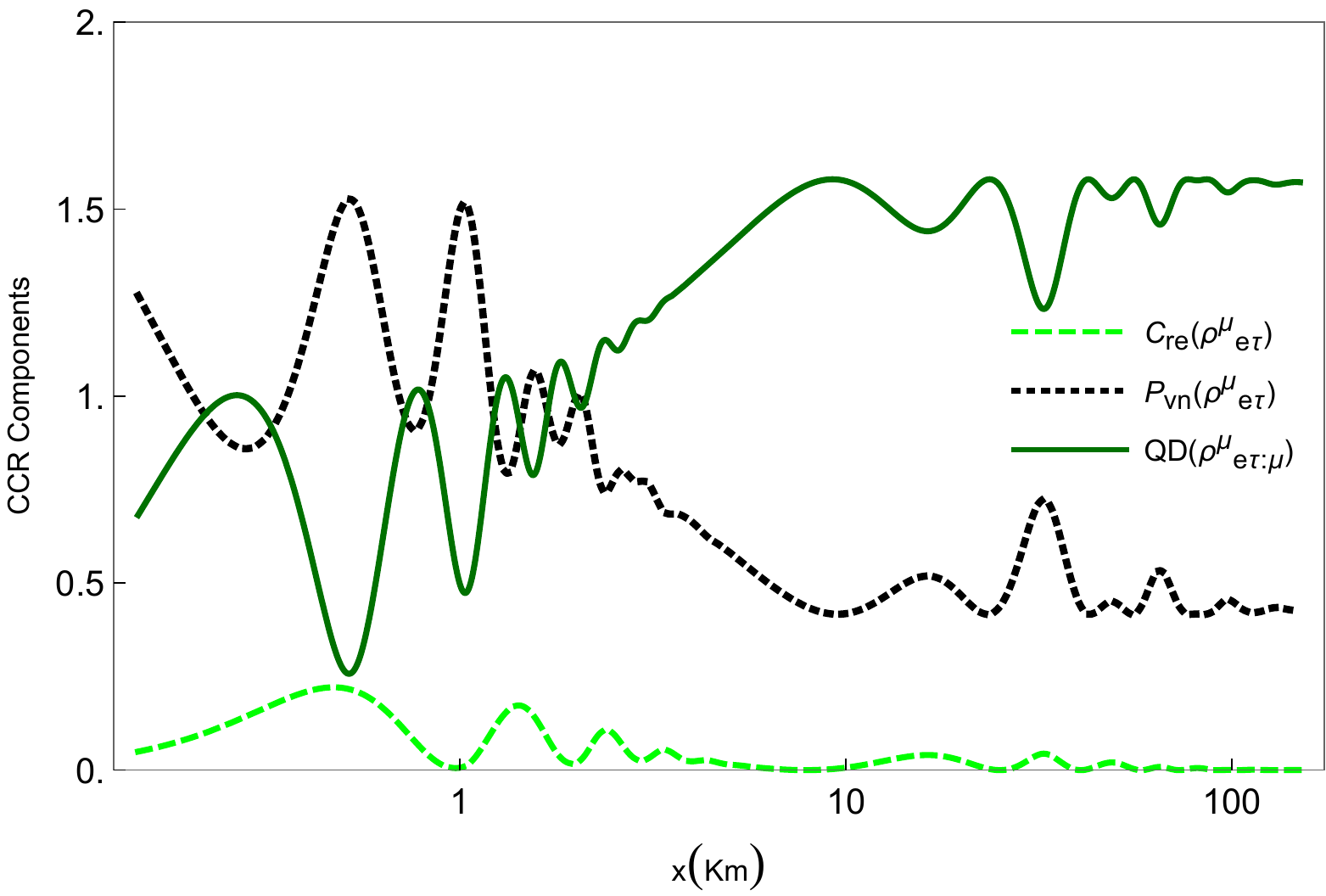}
}
\subfloat[]{
\includegraphics[width=5.5cm]{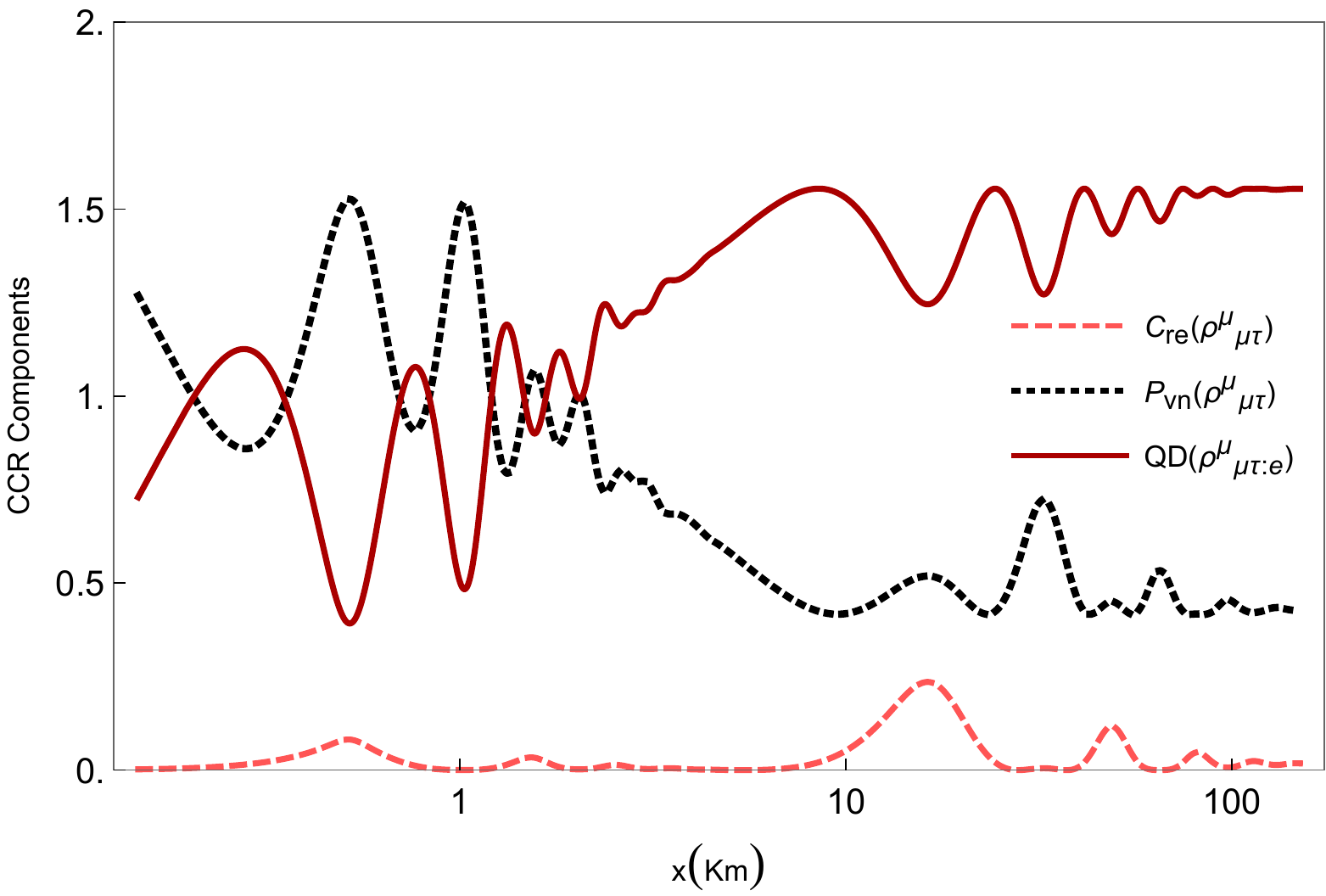}
}
\caption{CCR terms for two flavor subsystems $e\mu$ (a), $e\tau$ (b) and  $\mu\tau$ (c) as function of $x$, in the case of an initial muon neutrino state.}
\label{figure8}
\end{minipage}
\end{figure}

In Fig.\ref{figure8} are shown the CCR terms of Eq.(\ref{27}), for a neutrino system, as function of the distance, for the three possible two-flavor subsystems $e\mu$, $e\tau$ and $\mu\tau$. Again we observe the different behavior depending on the bipartite subsystem under consideration.

Quantum Discord again dominates the CCR at great distances. We also notice a trade-off among the ``local'' coherences, associated to $e\mu$, $e\tau$ and $\mu\tau$ subsystems, with the distance:  $C_{re}(\rho^{\mu}_{e\tau})$, which is initially the larger coherence, becomes close to zero at large $x$, where the maximum coherence is given by $C_{re}(\rho^{\mu}_{\mu\tau})$.


In general the level of coherence is much lower than in the initial electron neutrino case.   

\onecolumn

\section{Genuine tripartite contribution}

CCR we have considered in Eq.(\ref{6}, \ref{11}) only give information about \textit{bipartite} correlations between A and BC, without distinguishing bipartite and \textit{true} tripartite correlations between subsystems, although the trade-off of bipartite correlations are a clue of these last contributions. One could try to quantify the tripartite contribution by exploiting monogamy or polygamy relations.\\

\textbf{Monogamy and polygamy relations}:
Quantum correlations cannot be freely shared among many parties. The monogamy relations describe the limit of shareability of correlations in multi-partite systems \cite{mon0}--\cite{mon4}. By using the monogamy properties we can obtain significant information about the structure of multipartite quantum correlations.

Let us consider a bipartite quantum correlation measure Q applied to a quantum state $\rho_{ABC}$. This correlation measure is said \textit{monogamous} if it satisfies the relation:
\begin{equation}
Q_{A|BC}\ge Q_{AB}+Q_{AC},
\label{35}
\end{equation}
where $Q_{AB}$ and $Q_{AC}$ are the bipartite correlations between A-B and A-C, while $Q_{A|BC}$ is a measure quantifying the degree of correlation between subsystems A and BC.

Eq.(\ref{35}) tells us that the sum of correlations between A and each of the other parties B and C cannot exceed the correlations between A and BC. 
However, not all the correlation measures satisfy monogamous relations. Instead, some of them  satisfy the so-called polygamy relations \cite{pol1}--\cite{pol3}.
For example the von Neumann measure of entanglement satisfies:
\begin{equation}
S_{vn}(\rho_{A|BC})\le S_{vn}(\rho_{AB})+S_{vn}(\rho_{AC}).
\label{36}
\end{equation}
In contrast to monogamy relations, which give us an upper bound for the bipartite sheareability of entanglement in multipartite systems, polygamy relations provide a lower bound for distribution of bipartite entanglement. 

Polygamy relations can be exploited to extract the residual correlations \cite{Residual,Residual2}, which represent a collective property of the three single-partite subsystems, and can be linked to genuine tripartite. 

\subsection{Neutrino tripartite pure state}

In the case of a neutrino tri-partite pure state, the residual correlation is given by:

\begin{equation}
S_{vn}^{R}(\rho_{ABC})=S_{vn}(\rho_{A|BC})-(S_{vn}(\rho_{AB})+S_{vn}(\rho_{AC}))
\label{37}
\end{equation}

This could  permit us to distinguish the bipartite and tripartite contribution to CCR. 
From Eq. (\ref{37}) we obtain:
\begin{equation}
S_{vn}(\rho_{A|BC})=S_{vn}^{R}(\rho_{ABC})+S_{vn}(\rho_{AB})+S_{vn}(\rho_{AC}),
\label{38}
\end{equation}
and by replacing Eq.(\ref{38}) in Eq.(\ref{6}) we obtain the CCR:
\begin{equation}
P_{vn}(\rho_{A})+C_{re}(\rho_{A})+S_{vn}^{R}(\rho_{ABC})+S_{vn}(\rho_{AB})+S_{vn}(\rho_{AC})=\log_{2}d_{A}.
\label{39}
\end{equation}


We could think of exploiting polygamy relations to made this distinction also for CCR in Eq.(\ref{9}). However, Basso and Maziero demonstrated in \cite{Basso2020} that $C_{hs}^{nl}(\rho_{A|BC})=C_{hs}^{nl}(\rho_{A|B})+C_{hs}^{nl}(\rho_{A|C})$. This means that complementarity relations  as in Eq.(\ref{9}) for a single-partite subsystem are completed by its bipartite correlations with the other subsystem. Thus, in this case, we do not have a genuine contribution because $C_{hs}^{R}(\rho_{ABC})=0$. This fact shows that the two CCR before introduced  are not completely equivalent. Therefore, in the following we only consider the entropic form of CCR to analyze tripartite correlations in neutrino systems.

We consider the terms of Eq.(\ref{39}) for the subsystem $e$. By evaluating the eigenvalues of the reduced density matrices we obtain:
\begin{eqnarray}
P_{vn}(\rho^{\alpha}_{e})&=&1+(P_{\alpha \mu}+P_{\alpha \tau})\log_{2}(P_{\alpha \mu}+P_{\alpha \tau})+P_{\alpha e}\log_{2}P_{\alpha e},\\
S_{vn}(\rho^{\alpha}_{e\mu})&=&-(P_{\alpha e}+P_{\alpha \mu})\log_{2}(P_{\alpha e}+P_{\alpha \mu})-P_{\alpha \tau}\log_{2}P_{\alpha \tau},\\
S_{vn}(\rho^{\alpha}_{e\tau})&=&-(P_{\alpha e}+P_{\alpha \tau})\log_{2}(P_{\alpha e}+P_{\alpha \tau})-P_{\alpha \mu}\log_{2}P_{\alpha \mu},\\ \nonumber
S_{vn}^{R}(\rho^{\alpha}_{e\mu\tau})&=&-P_{\alpha e}\log_{2}P_{\alpha e}+P_{\alpha \mu}\log_{2}P_{\alpha \mu}+P_{\alpha \tau}\log_{2}P_{\alpha \tau}+(P_{\alpha e}+P_{\alpha \mu})\log_{2}(P_{\alpha e}+P_{\alpha \mu})\\ 
&+&(P_{\alpha e}+P_{\alpha \tau})\log_{2}(P_{\alpha e}+P_{\alpha \tau})-(P_{\alpha \mu}+P_{\alpha \tau})\log_{2}(P_{\alpha \mu}+P_{\alpha \tau}).
\end{eqnarray}
Similar expressions are valid for subsystems $\mu$, $\tau$, and we notice that the relative entropy of coherence vanishes for all three single-partite subsystems.

\begin{figure}[t]
\begin{minipage}{.45\textwidth}
\includegraphics[width=8cm]{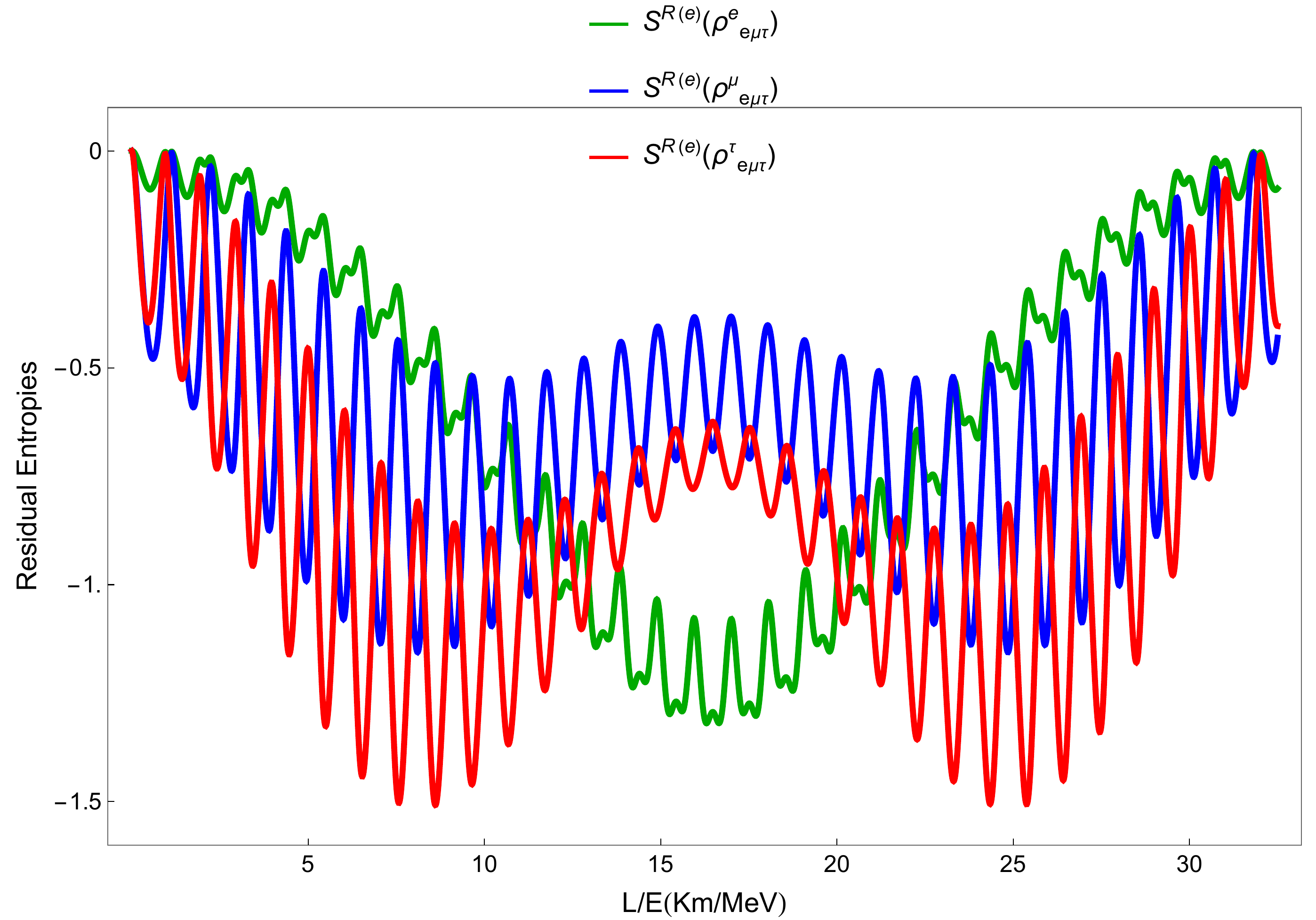}
\caption{ Residual terms for  subsystems $e$, $\mu$ and  $\tau$ as function of $L/E$ in the case of an initial electronic neutrino.}
\label{figure9}
\end{minipage}
\hspace{1.5cm}
\begin{minipage}{.45\textwidth}
\flushright
\includegraphics[width=8cm]{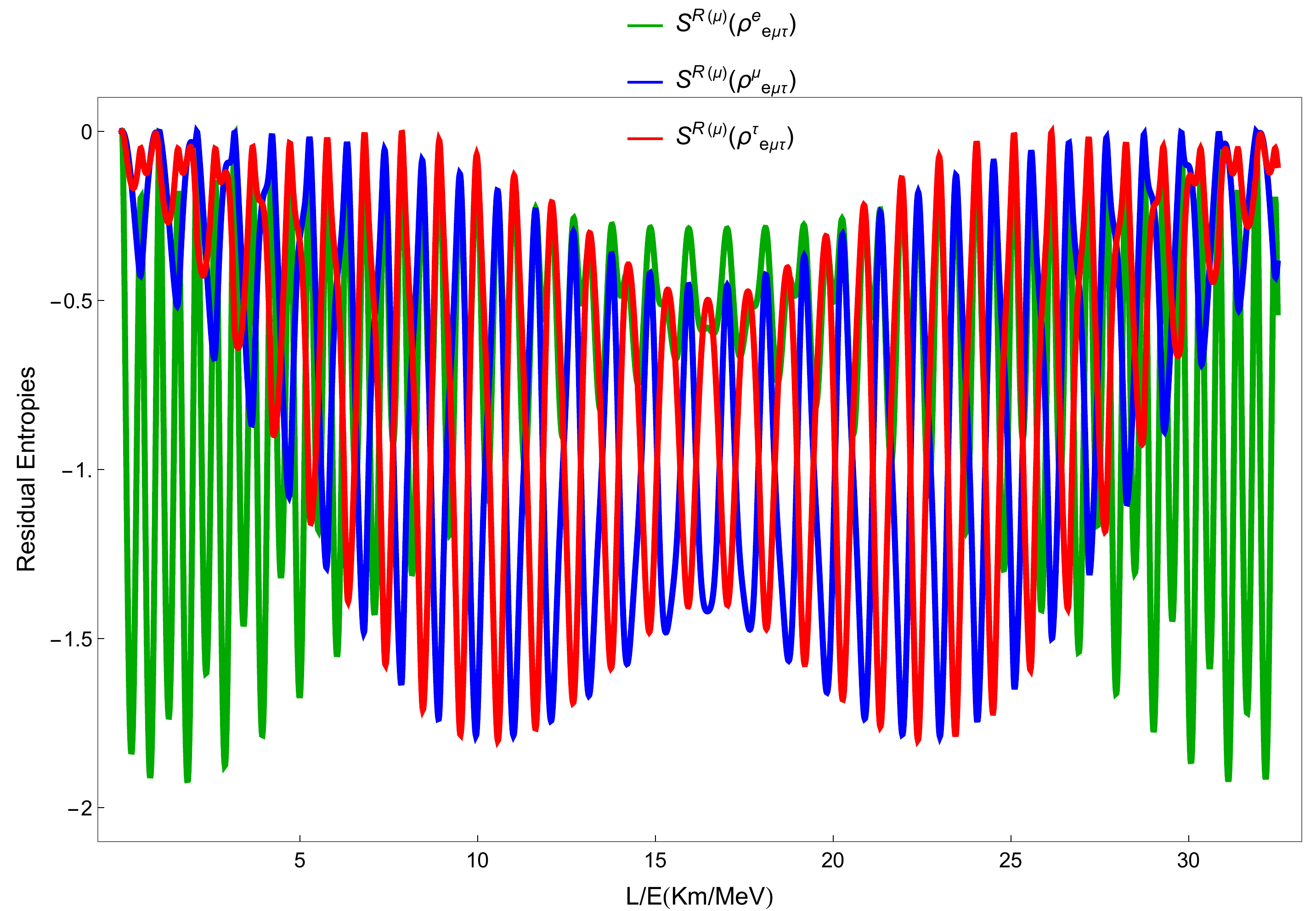}
\caption{Residual terms for  subsystems $e$, $\mu$ and  $\tau$ as function of $L/E$ in the case of an initial muonic neutrino.}
\label{figure10}
\end{minipage}
\end{figure}

In Fig.(\ref{figure9}) and Fig.(\ref{figure10}) we show the residual term of Eq.(\ref{39}) as function of the $L/E$, for an initial electronic and muonic neutrino system and for the three possible single-partite subsystems $e$, $\mu$ and $\tau$. Each partition considered show a different oscillatory behavior, i.e. $S_{vn}^{R}(\rho^{e}_{e\mu\tau})\ne S_{vn}^{R}(\rho^{\mu}_{e\mu\tau})\ne S_{vn}^{R}(\rho^{\tau}_{e\mu\tau})$. This does not allow us to consider the residual term  as a quantifier of  the genuine correlation shared among subsystems, because, if this were the case, the three residual correlations should coincide. In other words, the residual terms should be invariant under permutations of flavor modes.

\begin{figure}[h]
\begin{minipage}{.45\textwidth}
\includegraphics[width=8cm]{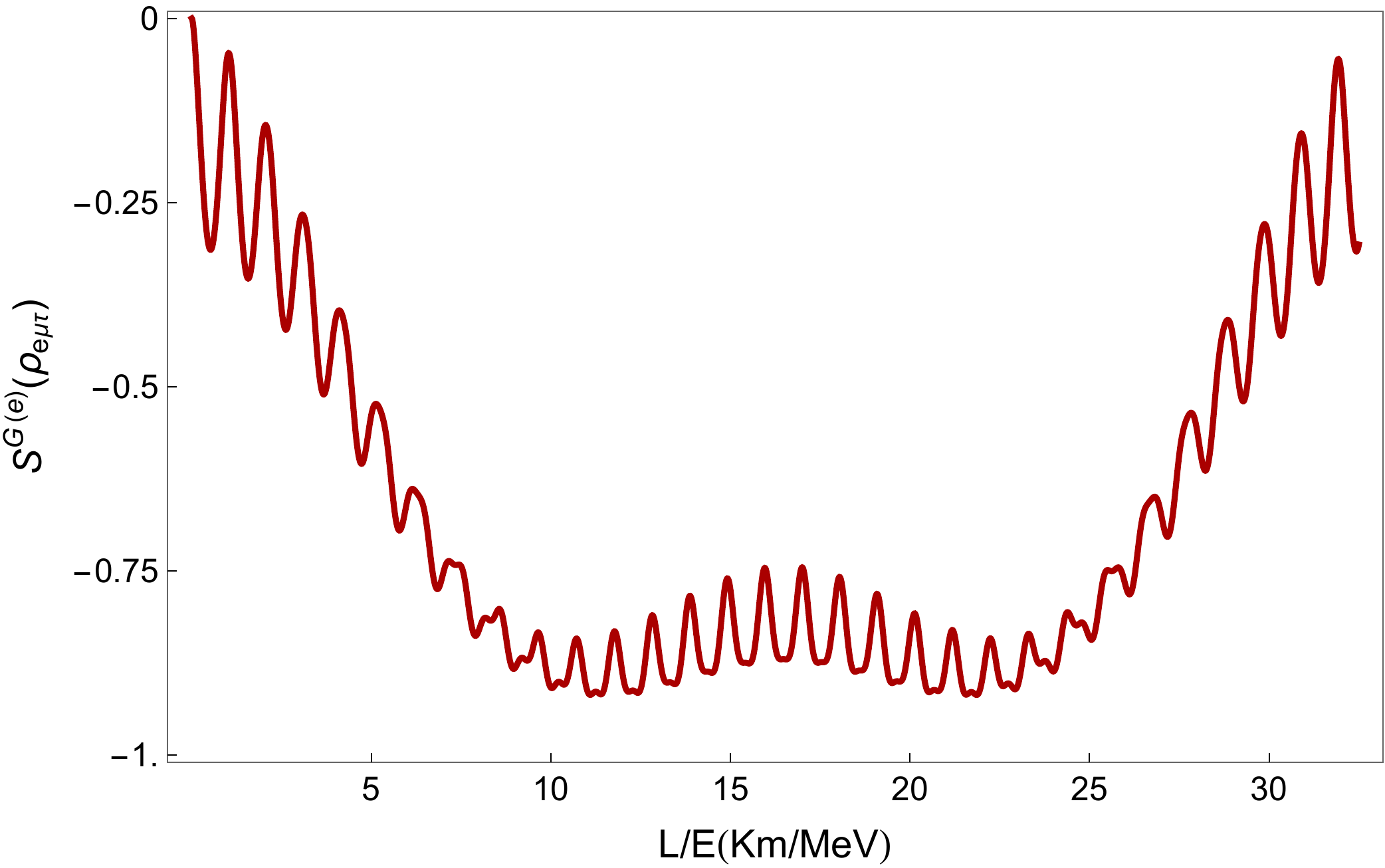}
\caption{Genuine tripartite correlations among subsystems $e$, $\mu$ and  $\tau$ as function of $L/E$ in the case of an initial electronic neutrino. }
\label{figure11}
\end{minipage}
\hspace{1.5cm}
\begin{minipage}{.45\textwidth}
\flushright
\includegraphics[width=8cm]{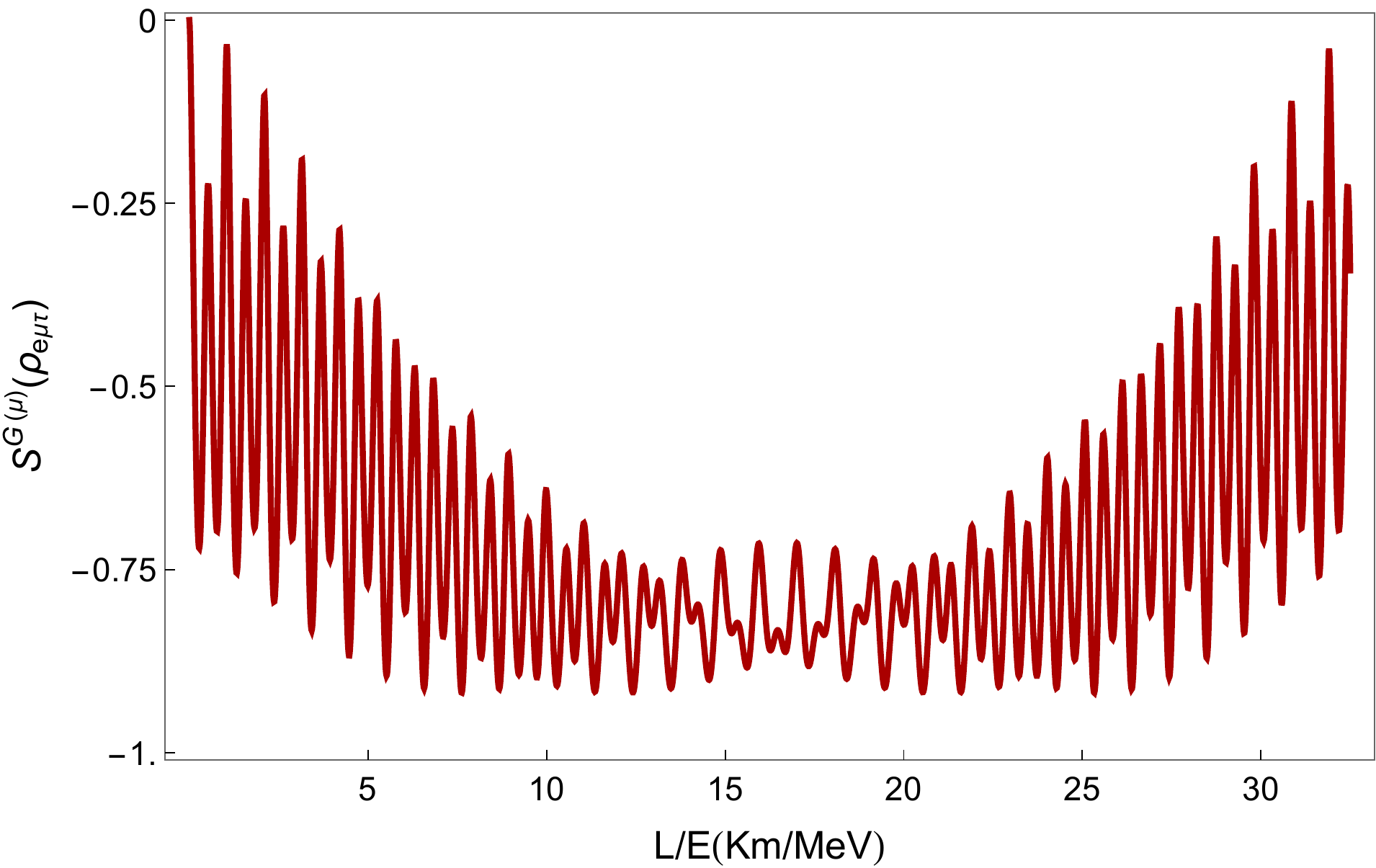}
\caption{Genuine tripartite correlations among subsystems $e$, $\mu$ and  $\tau$ as function of $L/E$ in the case of an initial muonic neutrino.}
\label{figure12}
\end{minipage}
\end{figure}

However, in the next section we quantify the tripartite genuine correlation by means of suitable averages.

%
%
%
%
%
%
%
%

\subsubsection{Tripartite entanglement}
Different from the bipartite case  \cite{Szalay,Walter}, the problem of multi-party correlation is not a simple issue and it is not clear at the moment whether it can be framed within the framework of the CCR. In \cite{B0}--\cite{gen}, the question is addressed in connection with the phenomena of particle mixing and oscillations. Several tripartite entanglement quantifiers have been proposed, and here we consider the average global entanglement and the averaged von-Neumann entropy. For a general measure of entanglement, it is possible to define a possible form of a genuine tripartite correlation as \cite{trijha,gen}:
\begin{equation}
E^{G}(\rho_{ABC})=\frac{E^{R}(\rho_{A})+E^{R}(\rho_{B})+E^{R}(\rho_{C})}{3},
\label{44}
\end{equation}
where $E^{R}(\rho_{x})=E(\rho_{x|yz})-E(\rho_{xy})-E(\rho_{xz})$, with $x,y,x=A,B,C$ and $x\ne y\ne z$.
Pursuing the same goal,the so-called average von Neumann entropy was defined in \cite{B0} as suitable weighted sum of  the von Neumann entropy associated with the all possible bipartitions of the system. In the case of tripartite system, the expression for this quantifier is given by:
\begin{equation}
\langle S^{3} \rangle= 
\begin{pmatrix}
3\\
2
\end{pmatrix}^{-1}
\!\!\!\!\!\!
\sum_{\substack{x=A,B,C\\ x\ne y \ne z}}\!\!\!\!\!\!
S^{(x-yz)}=\frac{1}{3}\left(S^{(A-BC)}+S^{(B-AC)}+S^{(C-AB)}\right),
\label{45}
\end{equation}
where $S^{(x-yz)}=S_{xyz}-S_{yz}$.
 By evaluating Eq.(\ref{44}) for a von Neumann measure of entanglement and making a comparison with the result one obtain from Eq.(\ref{45}), for the neutrino state under examination we find the following same expression for the genuine tripartite entanglement:
\begin{equation}
\begin{split}
S^{G(\alpha)}(\rho_{e\mu\tau})&=\frac{1}{3}\Big[P_{\alpha e}\log_{2}P_{\alpha e}+P_{\alpha\mu}\log_{2}P_{\alpha\mu}+P_{\alpha\tau}\log_{2}P_{\alpha\tau}+(P_{\alpha e}+P_{\alpha\mu})\log_{2}(P_{\alpha e}+P_{\alpha\mu})\\
&+(P_{\alpha e}+P_{\alpha\tau})\log_{2}(P_{\alpha e}+P_{\alpha \tau})+(P_{\alpha\mu}+P_{\alpha\tau})\log_{2}(P_{\alpha\mu}+P_{\alpha\tau})
\Big],
\end{split}
\label{46}
\end{equation}
which is invariant under permutation of flavor modes. In Fig. \ref{figure11} and Fig. \ref{figure12} we plot expression in Eq.(\ref{46})  for an electron and muon neutrino flavor state, respectively, as function of $L/E$.

\subsection{Neutrino tripartite mixed state}
A polygamy relation as in Eq.(\ref{36}) can be considered for quantum discord, from which it is possible to extract the residual tripartite discord:
\begin{equation}
QD^{R}(\rho_{ABC})=QD(\rho_{A|BC})-(QD(\rho_{AB})+QD(\rho_{AC})).
\label{47}
\end{equation}
Remembering that the sum of the two non-local terms of Eq.(\ref{28}) is equal to the quantum discord $QD(\rho_{A|BC})$, by means Eq.(\ref{47}) it is possible to rewrite the CCR for tripartite mixed states as:
\begin{equation}
    P_{vn}(\rho_{A})+C_{re}(\rho_{A})+QD^{R}(\rho_{ABC})+QD(\rho_{AB})+QD(\rho_{AC})=\log_{2}d_{A}.
    \label{48}
    \end{equation}

Starting with the density matrix in Eq.(\ref{29}), we evaluate the reduced density matrices for bi-partite and single-partite subsystems and the corresponding eigenvalues, it is possible to evaluate the terms of CCR for tripartite mixed state in Eq.(\ref{48}). 
In Appendix \ref{B}, we show the explicit expressions for the electronic subsystems.

\begin{figure}[t]
\begin{minipage}{.45\textwidth}
\includegraphics[width=8cm]{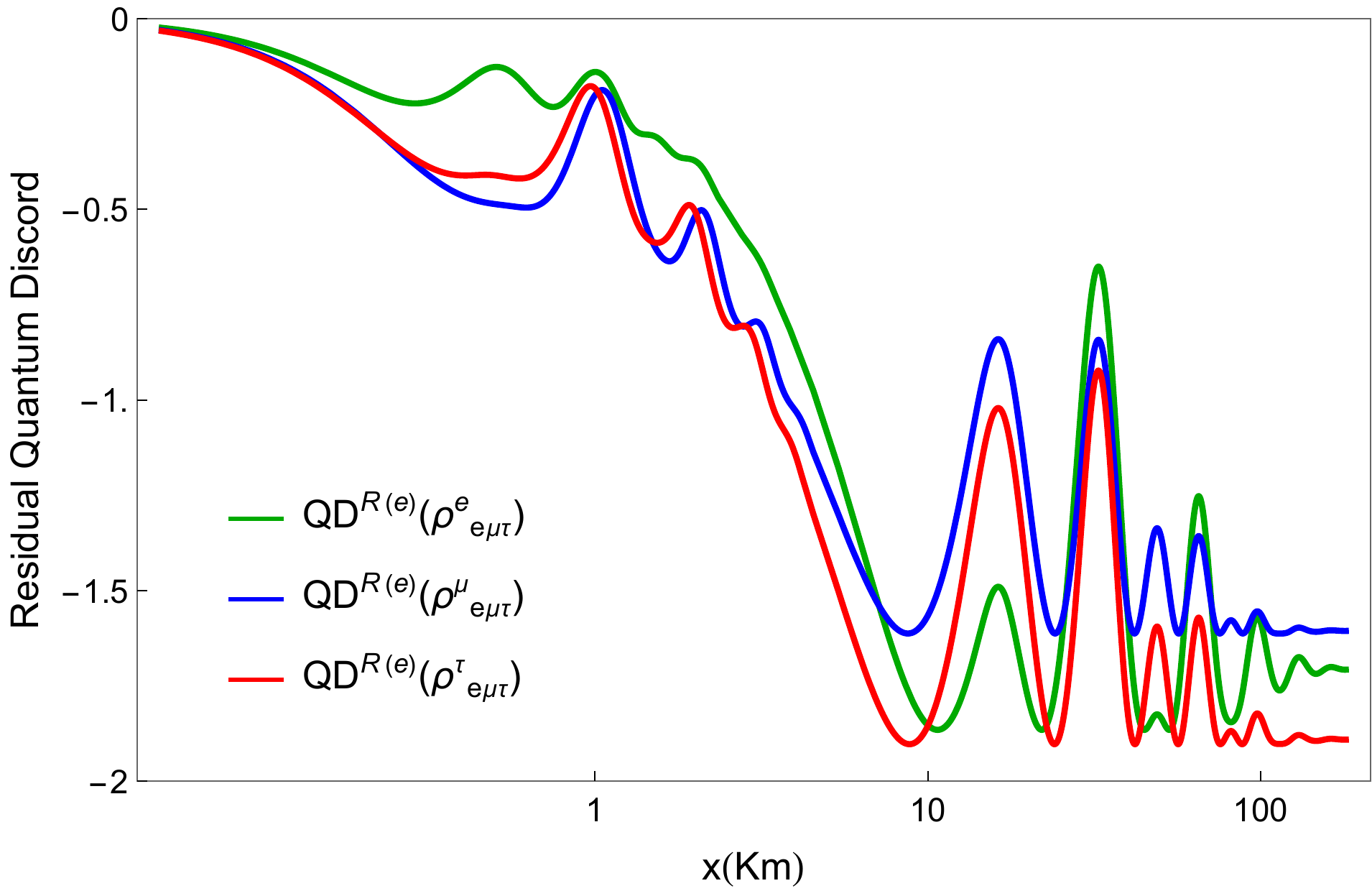}
\caption{ Residual discord for  subsystems $e$, $\mu$ and  $\tau$ as function of $x$ in the case of an initial electronic neutrino.}
\label{figure13}
\end{minipage}
\hspace{1.5cm}
\begin{minipage}{.45\textwidth}
\flushright
\includegraphics[width=8cm]{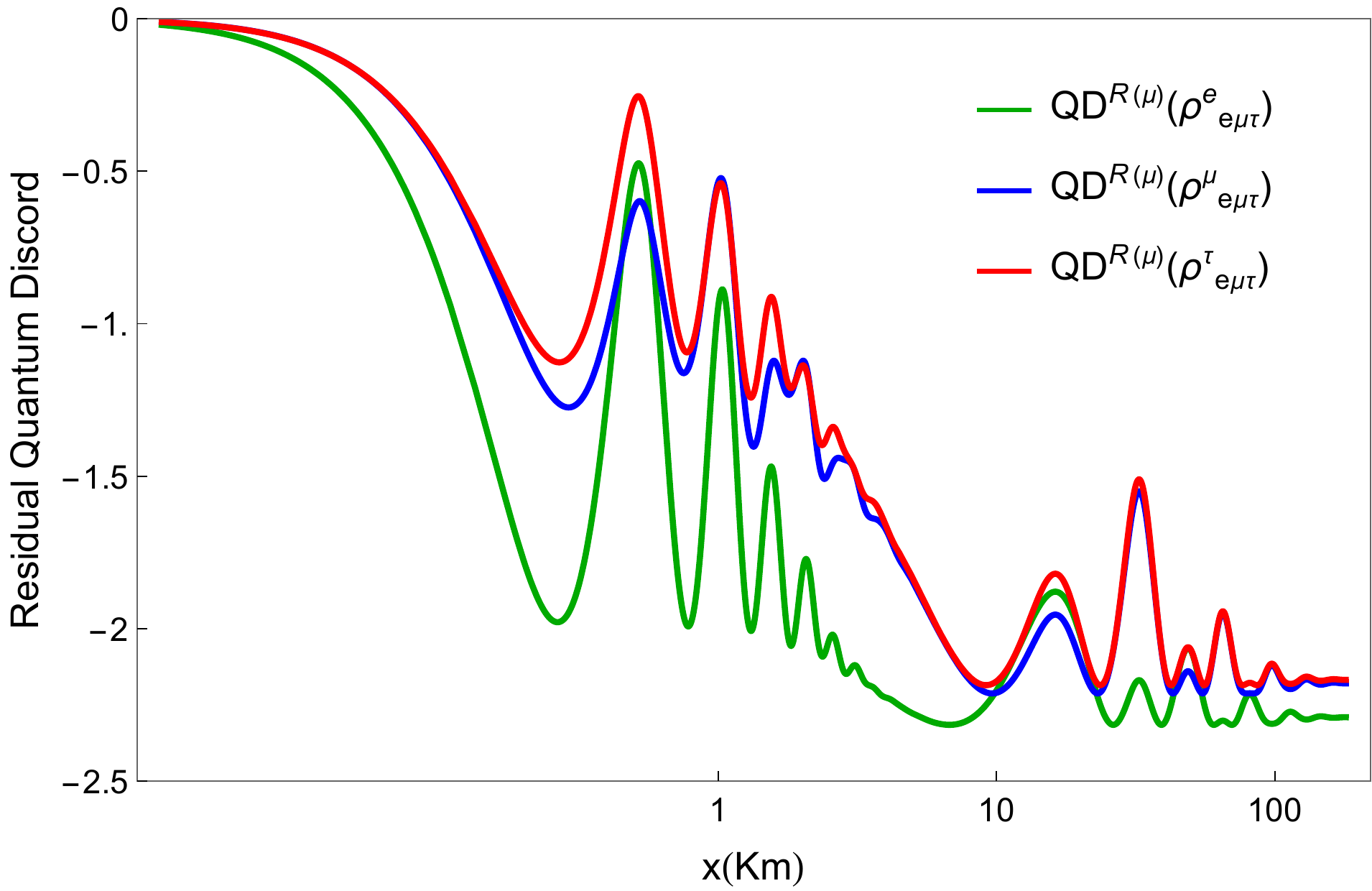}
\caption{Residual discord for  subsystems $e$, $\mu$ and  $\tau$ as function of $x$ in the case of an initial muonic neutrino.}
\label{figure14}
\end{minipage}
\end{figure}

In Fig.(\ref{figure13}) and Fig.(\ref{figure14})  the residual term of Eq.(\ref{48}) is shown, for an initial electronic and muonic neutrino system, respectively, as function of the distance $x$, for the three possible single-partite subsystems $e$, $\mu$ and $\tau$. The different expressions of the three residual discords do not allow  us to consider the residual terms  as a quantifier of the genuine tripartite discord shared among subsystems for the same reason discussed before: they are not invariant under permutation of the flavor index.

\subsubsection{Tripartite discord}
Analogously to the tripartite entangled that we have used in Section 4.1.1, we now analyze the behavior of the average of the three residual quantum discords as a genuine tripartite discord quantifier. It is given by:
\begin{equation}
QD^{G}(\rho_{ABC})=\frac{QD^{R}(\rho_{A})+QD^{R}(\rho_{B})+QD^{R}(\rho_{C})}{3}.
\label{53}
\end{equation}

We show the explicit expression in Eq.(\ref{53}) in Appendix \ref{B}

\begin{figure}[h]
\begin{minipage}{.45\textwidth}
\includegraphics[width=8cm]{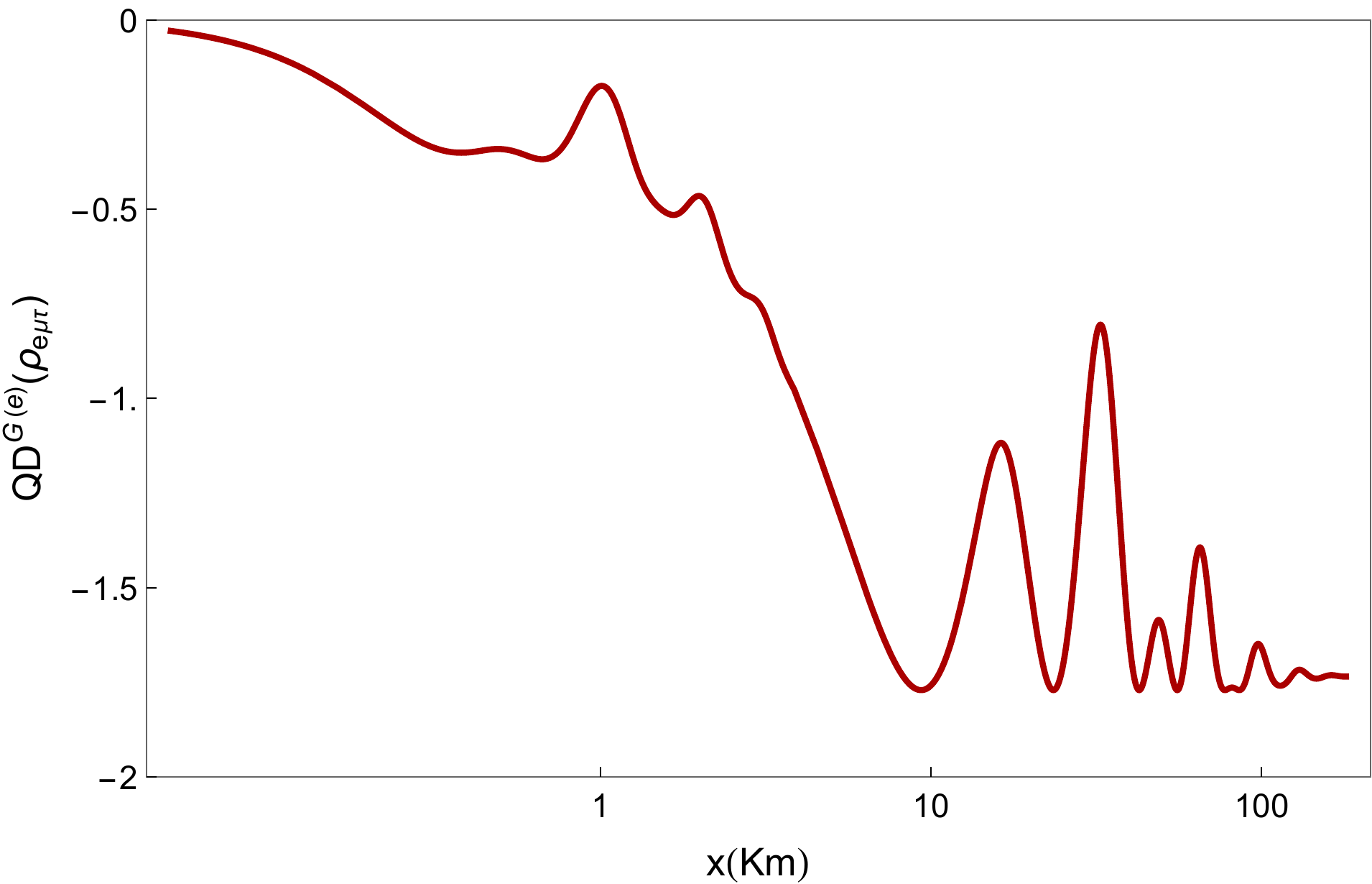}
\caption{Genuine tripartite quantum discord among subsystems $e$, $\mu$ and  $\tau$ as function of $x$ in the case of an initial electronic neutrino. }
\label{figure15}
\end{minipage}
\hspace{1.5cm}
\begin{minipage}{.45\textwidth}
\flushright
\includegraphics[width=8cm]{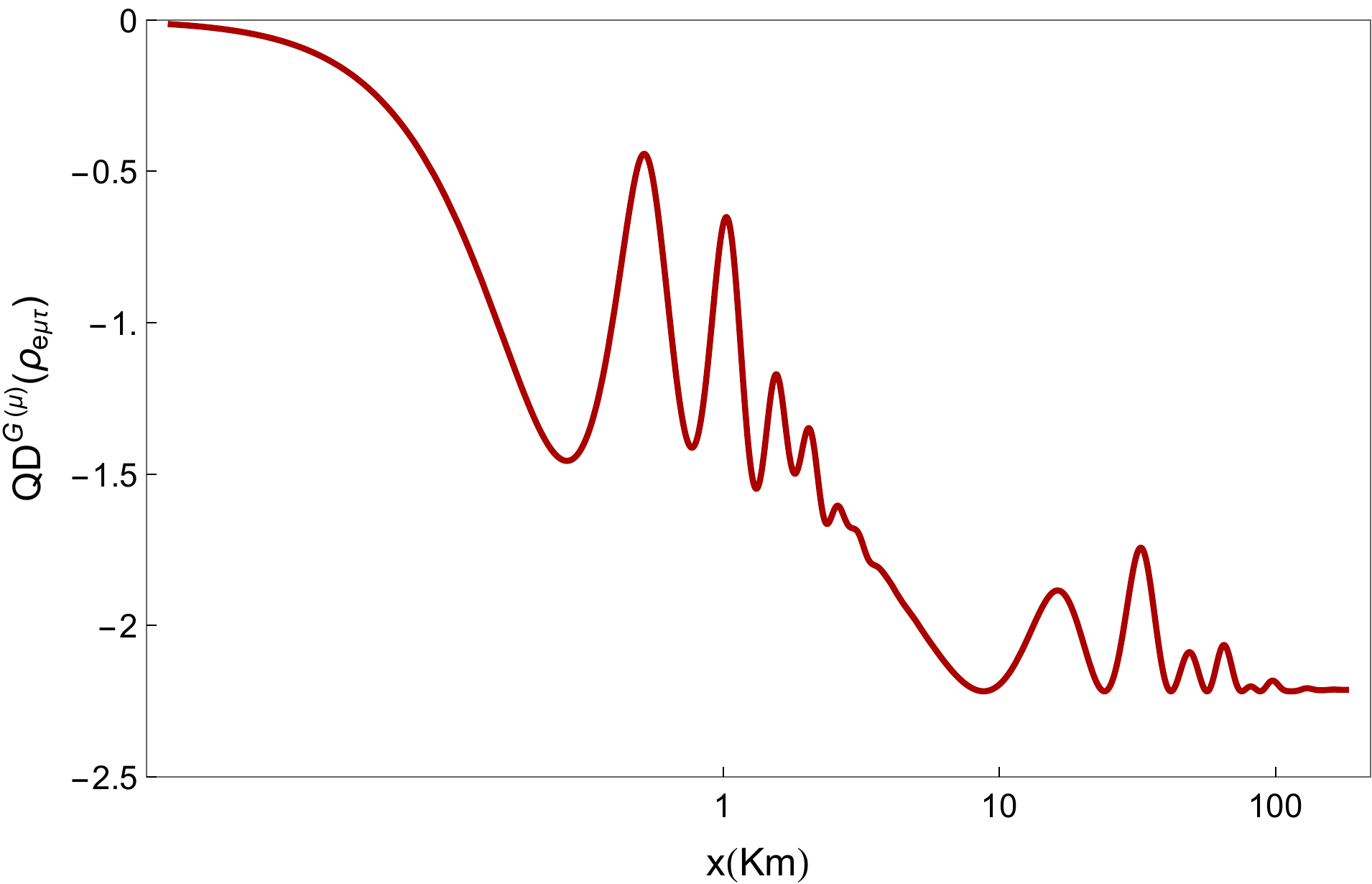}
\caption{Genuine tripartite quantum discord among subsystems $e$, $\mu$ and  $\tau$ as function of $x$ in the case of an initial muonic neutrino.}
\label{figure16}
\end{minipage}
\end{figure}

For the three-flavor neutrino state under examination, we can show  that the genuine quantum discord $QD^{G}(\rho^{\alpha}_{e\mu\tau})$ is invariant under permutation of flavor modes. In Fig.(\ref{figure15}) and Fig.(\ref{figure16}) we plot the genuine tripartite quantum discord  for an electron and muon neutrino flavor state, respectively, as function of $x$. We can note how $QD^{G}(\rho_{e\mu\tau})$ does not vanish at large distances, by denoting a persistent presence of true tripartite correlations among subsystems.

\section{Conclusions and outlook}
In this paper we have extended and completed previous investigations \cite{CCR2,CCR3piane} aimed to characterize quantum correlations involved in neutrino flavor oscillations via CCR.

We have studied in detail three-flavor neutrino oscillations within the wave-packet approach: CCR reveal a complex structure of correlations as function of the distance from the source and a trade-off among the different terms entering the complementarity relations. Our analysis confirms the persistence of quantum correlations, in particular true multipartite correlations, for wave-packet neutrinos at large distances \cite{Us,CCR2}.


A natural question arising in the three-flavor instance is about the possible presence of genuine tripartite correlations. Also it is interesting to ask if such contributions may be incorporated in the CCR. We find a positive answer for the first question: indeed, in the pure case the genuine tripartite correlation can be quantified both as the average of the three residual correlations associated to the three single-partite subsystems and as the average of the three bipartite correlations, and in both cases  the same result is obtained. Analogously, for mixed states we analyze the genuine tripartite contribution by using the concept of quantum discord. 

Regarding the second question, i.e. if we can include the genuine tripartite information in CCR, we have not obtained definite results. In fact, for the pure state case, by means of polygamy relations we attempted to extract a genuine tripartite contribution in the CCR, besides bipartite ones, by exploiting the definition of residual correlation. However, the three possible expressions for the residual correlations are not invariant for permutations of the flavors. A similar issue emerges in the mixed state instance.  

In the pure case, we find a genuine tripartite contribution by averaging the three residual correlations and by  suitable weighted sum of the von Neumann entropy
associated with the all possible bipartitions of the system, obtaining the same result in both cases. In the mixed case, we exploited the average of the residual correlations.

As an outlook, we plan to extend this work by including a non-zero CP-violating phase in the PNMS matrix. Another possible development is the inclusion of the spinorial nature of neutrinos and the consequent presence of chiral oscillations, in the line of Ref.\cite{BBZ23} where neutrino is an hyperentangled state. Finally, a further extension of the present analysis would be the formulation of the problem in the framework of QFT, by following the approach presented in Refs.\cite{BV1,BV2}.

\appendix

\section{Basics of neutrino oscillations theory in a plane wave approximation.}
\label{A}

Let us consider a neutrino state $\ket{\Psi(t)}$ al time $t$. It can be written in the flavor basis or in mass basis as \cite{BilPon,Giunti0}:
\begin{equation}
	\ket{\Psi(t)}=\sum_{\alpha}\nu_{\alpha}(t)\ket{\nu_{\alpha}}=\sum_{i}\nu_{i}(t)\ket{\nu_{i}},
\label{a2}
\end{equation}
where $\alpha=e,\mu,\tau$, $i=1,2,3$, $\ket{\nu_{\alpha}}$ and $\ket{\nu_{i}}$ are the flavor and mass eigenstates, respectively. The two representation are connected by an unitary matrix known as PMNS (Pontecorvo-Maki-Nakagawa-Sakat) mixing matrix, characterized by three mixing angles $(\theta_{12},\theta_{13},\theta_{23})$  and a charge conjugation and parity (CP) violating phase:
\begin{equation}
	U(\theta_{ij},\delta)=
	\begin{pmatrix}
		c_{12}c_{13} & s_{12}c_{13} & s_{23}e^{-i\delta} \\
		-s_{12}c_{23} - c_{12}s_{23}s_{13}e^{i\delta}& c_{12}c_{23}-s_{12}s_{23}s_{13}e^{i\delta} & s_{23}c_{13}\\
		s_{13}s_{23} - c_{12}c_{23}s_{13}e^{i\delta}& -c_{12}s_{23}-s_{12}c_{23}s_{13}e^{i\delta} & c_{23}c_{13}
	\end{pmatrix}
	\label{a3}
\end{equation}
where $c_{ij}=\cos \theta_{ij}$, $s_{ij}=\sin \theta_{ij}$.\\
$\ket{\nu_{i}}$ are eigenstates of the Hamiltonian $H$ with energy $E_{i}=\sqrt{\vec{p}^{2}+m_{i}^{2}}$: $H\ket{\nu_{i}}=E_{i}\ket{\nu_{i}}$. By resolving the Schr\"{o}dinger equation for the mass eigenstates, given by:
\begin{equation}
	i\frac{d}{dt}\ket{\nu_{i}(t)}=H\ket{\nu_{i(t)}},
	\label{a4}
\end{equation}
we obtain:
\begin{equation}
	\ket{\nu_{i}(t)}=e^{-iE_{i}t}\ket{\nu_{i}},
	\label{a5}
\end{equation}
which tells us that the mass eigenstates evolve in time as plane waves.
By expressing the time evolution of a flavor state $\ket{\nu_{\alpha}}$ at $t=0$ in terms of the mass one through the mixing matrix elements:
\begin{equation}
	\ket{\nu_{\alpha}(t)}=\sum_{i}U_{\alpha i}^{*}\ket{\nu_{i}(t)}=\sum_{i}U_{\alpha i}^{*}e^{-iE_{i}t}\ket{\nu_{i}},
	\label{a6}
\end{equation}
By remembering that we can write the mass eigenstate in terms of flavor ones as $\ket{\nu_{i}}=\sum_{\beta}U_{\beta i}\ket{\nu_{\beta}}$, Eq. (\ref{a6}) becomes:
\begin{equation}
	\ket{\nu_{\alpha}(t)}=\sum_{\beta}\sum_{i}U_{\alpha i}^{*}e^{-iE_{i}t}U_{\beta i}\ket{\nu_{\beta}}=\sum_{\beta}\tilde{U}_{\alpha\beta}(t)\ket{\nu_{\beta}},
	\label{a7}
\end{equation}
where $\tilde{U}_{\alpha\beta}(t)=\sum_{i}U_{\alpha i}^{*}e^{-iE_{i}t}U_{\beta i}$ is the transition amplitude from flavor $\alpha$ to flavor $\beta$.\\
Thus, while at $t=0$ $\ket{\nu_{\alpha}}$ is a pure flavor state, at $t>0$ it becomes a superposition of different flavor states.\\
It is simple to obtain the transition probability as $|\bra{\nu_{\beta}}\ket{\nu_{\alpha}(t)}|^{2}=|\tilde{U}_{\alpha\beta}|^{2}$:
\begin{equation}
	P_{\alpha\beta}=\sum_{ik}U_{\alpha i}^{*}U_{\beta i}U_{\alpha k}U_{\beta k}^{*}e^{-i(E_{i}-E_{k})t}.
	\label{a8}
\end{equation}
From the moment that neutrinos possess a velocity nearly of speed of light, we can consider the ultra-relativistic approximation:
\begin{equation}
	E_{i}\simeq E+\frac{m_{i}^{2}}{2E},
	\label{a9}
\end{equation}
 where $E$ is the energy of neutrino in the limit of zero mass. $E_{i}$ and $m_{i}$ are the energy and the mass of the mass eigenstate $\ket{\nu_{i}}$, respectively. By using this approximation, we can write:
 \[
 E_{i}-E_{k}\simeq \frac{\Delta m_{ik}^{2}}{2E}, \hspace{1cm} \Delta m_{ik}^{2}=m_{i}^{2}-m_{k}^{2}.
 \]
Again, in the limit under consideration, we can approximate $L=t$ and write the transition probability as:
\begin{equation}
	P_{\alpha\beta}(L,E)=\sum_{ik}U_{\alpha i}^{*}U_{\beta i}U_{\alpha k}U_{\beta K}^{*}e^{-i\frac{\Delta m_{ik}^{2}L}{2E}}.
	\label{a10}
\end{equation}
Oscillations are possible only if neutrino masses are non-zero and $L>0$. In fact, if the phase goes to zero, due to the unitary relation $UU^{\dagger}=1 \leftrightarrow \sum_{i}U_{\alpha i}U_{\beta i}^{*}=\delta_{\alpha\beta}$, we have:
\[P_{\alpha\beta}=
\begin{cases}
	1\hspace{1cm} if \hspace{0.5cm} \alpha=\beta\\
	0\hspace{1cm}if \hspace{0.5cm}\alpha\ne\beta
	\end{cases}\]

This derivation assumes that neutrinos have definite energy and  permits to consider plane waves infinitely extended in space and time. But this is clearly unrealistic from the moment that experiments take place in finite time and space. For a more sophisticated  characterization of neutrino oscillations it is convenient to use a wave packet approach.

\subsection{Wave packet approach}

We briefly review the theory of neutrino oscillation in the wave packet approach \cite{Giunti1,Giunti2}.

Let us consider a  neutrino with definite flavor $\alpha\, (\alpha=e,\mu,\tau)$, that propagates along x axis. It can be described by:

\begin{equation}
\ket{\nu_{\alpha}(t)}=\sum_{j=1}^3U^{*}_{\alpha j}\psi_{j}(x,t)\ket{\nu_{j}},
\label{n1}
\end{equation}
where $U_{\alpha j}$ denotes the elements of the PMNS mixing matrix and $\psi_{j}(x,t)$ is the wave function of the mass eigenstates $\ket{\nu_{j}}$ with mass $m_{j}$. The wave function can be written as:
\begin{equation}
\psi_{j}(x,t)=\frac{1}{\sqrt{2\pi}}\int dp\hspace{0.1cm} \psi_{j}(p)e^{ipx-iE_{j}(p)t},
\label{n2}
\end{equation}
with $\psi_{j}(p)=(2\pi\sigma_{p}^{2})^{-1/4}e^{-\frac{1}{4\sigma_{p}^{2}}(p-p_{j})^{2}}$, where $p_{j}$ is the average momentum and $\sigma_{p}$ is the momentum uncertainty determined by the production process. $E_{j}(p)=\sqrt{p^{2}+m_{j}^{2}}$ is the energy.
By assuming that the Gaussian momentum distribution is strongly peaked around $p_{j}$, i.e. $\sigma_{p}\ll E^{2}_{j}(p_{j})/m_{j}$, it is possible to approximate the energy as $E_{j}(p)\simeq E_{j}+v_{j}(p-p_{j})$, where $E_{j}=\sqrt{p_{j}^{2}+m_{j}^{2}}$ is the average energy and $v_{j}$ is the group velocity of the wave packet of the massive neutrino $\nu_{j}$.  Using this approximation an integration over $p$ of Eq.(\ref{n2}) can be performed, obtaining:
\small
\begin{equation}
\psi_{j}(x,t)=(2\pi\sigma_{x}^{2})^{-\frac{1}{4}}\exp\left[-iE_{j}t+ip_{j}x-\frac{(x-v_{j}t)^{2}}{4\sigma_{x}^{2}}\right],
\label{n3}
\end{equation}
\normalsize
where $\sigma_{x}=\frac{1}{\sigma_{p}}$ is the spatial width of the wave packet.
At this point, by substituting Eq.(\ref{n3}) in Eq.(\ref{n1}) it is possible to obtain the density matrix $\rho_{\alpha}(x,t)=\ket{\nu_{\alpha}(x,t)}\bra{\nu_{\alpha}(x,t)}$ describing the neutrino oscillations in space and time.
In the case of ultra-relativistic neutrinos, it is useful to consider the following approximations: $E_{j}\simeq E+\xi\frac{m_{j}^{2}}{2E}$, where $E$ is the neutrino energy in the limit of zero mass and $\xi$ is a dimensionless quantity that depends on the characteristics of the production process, $p_{j}\simeq E-(1-\xi)\frac{m_{j}^{2}}{2E}$ and $v_{j}\simeq 1-\frac{m_{j}^{2}}{2E_{j}}$.

 Although in laboratory experiments it is possible to measure neutrino oscillations in time through the measurements of both the production and detection process, due to the long time exposure of the detectors it is convenient to consider an average in time of the density matrix operator and it can be obtained by a Gaussian time integration of $\rho_{\alpha}(x,t)$:
\begin{equation}
\rho_{\alpha}(x)=\sum_{k,j}U_{\alpha k}U_{\alpha j}^{*}f_{jk}(x)\ket{\nu_{j}}\bra{\nu_{k}},
\label{n4}
\end{equation}
with:
\begin{equation}
f_{jk}(x)=\exp\left[-i\frac{\Delta m_{jk}^{2}x}{2E}-\left(\frac{\Delta m_{ij}^{2}x}{4\sqrt{2}E^{2}\sigma_{x}}\right)^{2}\right].
\label{n5}
\end{equation}
Here, $\Delta m_{jk}^{2}=m_{j}^{2}-m_{k}^{2}$.
 It is convenient to express $\rho_{\alpha}(x)$ in terms of flavor eigenstates by establishing the identification $\ket{\nu_{\alpha}}=\ket{\delta_{\alpha e}}_{e}\ket{\delta_{\alpha \mu}}_{\mu}\ket{\delta_{\alpha \tau}}_{\tau}$ \cite{Blas0}. 

Using $\ket{\nu_{i}}=\sum_{\alpha}U_{\alpha i}\ket{\nu_{\alpha}}$, we can write:
\begin{equation}
\rho_{\alpha}(x)=\sum_{\beta\gamma}F^{\alpha}_{\beta\gamma}\ket{\delta_{\beta e}\delta_{\beta\mu}\delta_{\beta\tau}}\bra{\delta_{\gamma e}\delta_{\gamma\mu}\delta_{\gamma\tau}},
\label{n6}
\end{equation}
where:
\begin{equation}
F^{\alpha}_{\beta\gamma}=\sum_{kj}U^{*}_{\alpha j}U_{\alpha k}f_{jk}(x)U_{\beta j}U_{\gamma k}^{*},
\label{n7}
\end{equation}
with $k,j=1,2,3$ and $\beta,\gamma=e,\mu,\tau$.

The transition probability for the neutrino described by $\rho_{\alpha}(x)$ to be in the flavor $\eta$ at position $x$ is given by:
\begin{equation}
P_{\alpha\rightarrow\eta}(x)=\bra{\nu_{\eta}}\rho_{\alpha}(x)\ket{\nu_{\eta}}=F^{\alpha}_{\eta\eta}(x).
\label{n8}
\end{equation}

It is worth to notice that density matrices as in Eq.(\ref{n4}) represent mixed states.

\section{Explicit expressions for tri-partite mixed neutrino states}
\label{B}
\begin{itemize}
\item Explicit expressions of  CCR terms of Eq.(\ref{27}) for the subsystem $e \mu$:
\end{itemize}

\begin{equation}
P_{vn}(\rho_{e\mu}^{\alpha})=\log_{2}(d_{e}d_{\mu})-S_{vn}(\rho_{e\mu_{diag}})=
2+F^{\alpha}_{ee}\log_{2}F^{\alpha}_{ee}+F^{\alpha}_{\mu\mu}\log_{2}F^{\alpha}_{\mu\mu}+F^{\alpha}_{\tau\tau}\log_{2}F^{\alpha}_{\tau\tau}.
\label{30}
\end{equation}

\vspace{0.01cm}

\begin{eqnarray}\nonumber
&C_{re}(\rho^{\alpha}_{e\mu})=S_{vn}(\rho^{\alpha}_{e\mu_{diag}})-S_{vn}(\rho_{e\mu}^{\alpha})=
-F^{\alpha}_{ee}\log_{2}F^{\alpha}_{ee}-F^{\alpha}_{\mu\mu}\log_{2}F^{\alpha}_{\mu\mu}+\\
&\frac{1}{2}\left(F^{\alpha}_{ee}+F^{\alpha}_{\mu\mu}-\sqrt{(F_{ee}^{\alpha}-F^{\alpha}_{\mu\mu})^{2}+4F^{\alpha}_{e\mu}F^{\alpha}_{\mu e}}\right)
\log_{2}\left[\frac{1}{2}\left(F^{\alpha}_{ee}+F^{\alpha}_{\mu\mu}-\sqrt{(F_{ee}^{\alpha}-F^{\alpha}_{\mu\mu})^{2}+4F^{\alpha}_{e\mu}F^{\alpha}_{\mu e}}\right)\right]+\nonumber\\
&\frac{1}{2}\left(F^{\alpha}_{ee}+F^{\alpha}_{\mu\mu}+\sqrt{(F_{ee}^{\alpha}-F^{\alpha}_{\mu\mu})^{2}+4F^{\alpha}_{e\mu}F^{\alpha}_{\mu e}}\right)
\log_{2}\left[\frac{1}{2}\left(F^{\alpha}_{ee}+F^{\alpha}_{\mu\mu}+\sqrt{(F_{ee}^{\alpha}-F^{\alpha}_{\mu\mu})^{2}+4F^{\alpha}_{e\mu}F^{\alpha}_{\mu e}}\right)\right]
\label{31}
\end{eqnarray}

\vspace{0.01cm}

\begin{eqnarray}\nonumber
&I_{e\mu:\tau}(\rho^{\alpha}_{e\mu\tau})=S_{vn}(\rho^{\alpha}_{e\mu})+S_{vn}(\rho^{\alpha}_{\tau})-S_{vn}(\rho^{\alpha}_{e\mu\tau})=-(F^{\alpha}_{ee}+F^{\alpha}_{\mu\mu})\log_{2}[F^{\alpha}_{ee}+F^{\alpha}_{\mu\mu}]-2F^{\alpha}_{\tau\tau}\log_{2}F^{\alpha}_{\tau\tau}-\\
&\frac{1}{2}\left(F^{\alpha}_{ee}+F^{\alpha}_{\mu\mu}-\sqrt{(F_{ee}^{\alpha}-F^{\alpha}_{\mu\mu})^{2}+4F^{\alpha}_{e\mu}F^{\alpha}_{\mu e}}\right)\log_{2}\left[\frac{1}{2}\left(F^{\alpha}_{ee}+F^{\alpha}_{\mu\mu}-\sqrt{(F_{ee}^{\alpha}-F^{\alpha}_{\mu\mu})^{2}+4F^{\alpha}_{e\mu}F^{\alpha}_{\mu e}}\right)\right]+\nonumber\\
&\frac{1}{2}\left(F^{\alpha}_{ee}+F^{\alpha}_{\mu\mu}+\sqrt{(F_{ee}^{\alpha}-F^{\alpha}_{\mu\mu})^{2}+4F^{\alpha}_{e\mu}F^{\alpha}_{\mu e}}\right)\log_{2}\left[\frac{1}{2}\left(F^{\alpha}_{ee}+F^{\alpha}_{\mu\mu}+\sqrt{(F_{ee}^{\alpha}-F^{\alpha}_{\mu\mu})^{2}+4F^{\alpha}_{e\mu}F^{\alpha}_{\mu e}}\right)\right]+\nonumber\\
&(F^{\alpha}_{ee}+F^{\alpha}_{\mu\mu}+F^{\alpha}_{\tau\tau})\log_{2}[F^{\alpha}_{ee}+F^{\alpha}_{\mu\mu}+F^{\alpha}_{\tau\tau}]
\label{32}
\end{eqnarray}

\vspace{0.01cm}

\begin{eqnarray}\nonumber
&S_{e\mu|\tau}(\rho^{\alpha}_{e\mu\tau})=S_{vn}(\rho^{\alpha}_{e\mu\tau})-S_{vn}(\rho_{\tau}^{\alpha})=(F^{\alpha}_{ee}+F^{\alpha}_{\mu\mu})\log_{2}[F^{\alpha}_{ee}+F^{\alpha}_{\mu\mu}]+F^{\alpha}_{\tau\tau}\log_{2}F^{\alpha}_{\tau\tau}-\\
&(F^{\alpha}_{ee}+F^{\alpha}_{\mu\mu}+F^{\alpha}_{\tau\tau})\log_{2}[F^{\alpha}_{ee}+F^{\alpha}_{\mu\mu}+F^{\alpha}_{\tau\tau}].
\label{33}
\end{eqnarray}

By adding all these terms, Eq.(\ref{27}) is verified. 

Similar expressions have been obtained for subsystems $e\tau$ and $\mu\tau$.

\begin{itemize}
\item Explicit expressions of the terms of Eq.(\ref{48}) for the subsystem $e$:
\end{itemize}

\begin{eqnarray}
    P_{vn}(\rho^{\alpha}_{e})&=&1+F^{\alpha}_{ee}\log_{2}F^{\alpha}_{ee}+(F^{\alpha}_{\mu\mu}+F^{\alpha}_{\tau\tau})\log_{2}(F^{\alpha}_{\mu\mu}+F^{\alpha}_{\tau\tau})\\ \nonumber
    QD(\rho^{\alpha}_{e\mu}),&=&- F^{\alpha}_{\tau\tau}\log_{2}F^{\alpha}_{\tau\tau}\\ \nonumber
    &-&\frac{1}{2}(F^{\alpha}_{ee}+F^{\alpha}_{\mu\mu}-\sqrt{(F^{\alpha}_{ee}-F^{\alpha}_{\mu\mu})^{2}+4F^{\alpha}_{e\mu}F^{\alpha}_{\mu e}})\log_{2}[\frac{1}{2}(F^{\alpha}_{ee}+F^{\alpha}_{\mu\mu}-\sqrt{(F^{\alpha}_{ee}-F^{\alpha}_{\mu\mu})^{2}+4F^{\alpha}_{e\mu}F^{\alpha}_{\mu e}})]\\
    &-&\frac{1}{2}(F^{\alpha}_{ee}+F^{\alpha}_{\mu\mu}+\sqrt{(F^{\alpha}_{ee}-F^{\alpha}_{\mu\mu})^{2}+4F^{\alpha}_{e\mu}F^{\alpha}_{\mu e}})\log_{2}[\frac{1}{2}(F^{\alpha}_{ee}+F^{\alpha}_{\mu\mu}+\sqrt{(F^{\alpha}_{ee}-F^{\alpha}_{\mu\mu})^{2}+4F^{\alpha}_{e\mu}F^{\alpha}_{\mu e}})], \\ \nonumber
     QD(\rho^{\alpha}_{e\tau}),&=&- F^{\alpha}_{\mu\mu}\log_{2}F^{\alpha}_{\mu\mu}\\ \nonumber
    &-&\frac{1}{2}(F^{\alpha}_{ee}+F^{\alpha}_{\tau\tau}-\sqrt{(F^{\alpha}_{ee}-F^{\alpha}_{\tau\tau})^{2}+4F^{\alpha}_{e\tau}F^{\alpha}_{\tau e}})\log_{2}[\frac{1}{2}(F^{\alpha}_{ee}+F^{\alpha}_{\tau\tau}-\sqrt{(F^{\alpha}_{ee}-F^{\alpha}_{\tau\tau})^{2}+4F^{\alpha}_{e\tau}F^{\alpha}_{\tau e}})]\\
    &-&\frac{1}{2}(F^{\alpha}_{ee}+F^{\alpha}_{\tau\tau}+\sqrt{(F^{\alpha}_{ee}-F^{\alpha}_{\tau\tau})^{2}+4F^{\alpha}_{e\tau}F^{\alpha}_{\tau e}})\log_{2}[\frac{1}{2}(F^{\alpha}_{ee}+F^{\alpha}_{\tau\tau}+\sqrt{(F^{\alpha}_{ee}-F^{\alpha}_{\tau\tau})^{2}+4F^{\alpha}_{e\tau}F^{\alpha}_{\tau e}})], \\ \nonumber
    QD^{R}(\rho^{\alpha}_{e\mu\tau})&=&-F^{\alpha}_{ee}\log_{2}F^{\alpha}_{ee}+ F^{\alpha}_{\mu\mu}\log_{2}F^{\alpha}_{\mu\mu}+F^{\alpha}_{\tau\tau}\log_{2}F^{\alpha}_{\tau\tau}-(F^{\alpha}_{\mu\mu}+F^{\alpha}_{\tau\tau})\log_{2}(F^{\alpha}_{\mu\mu}+F^{\alpha}_{\tau\tau})\\ \nonumber
    &+&\frac{1}{2}(F^{\alpha}_{ee}+F^{\alpha}_{\mu\mu}-\sqrt{(F^{\alpha}_{ee}-F^{\alpha}_{\mu\mu})^{2}+4F^{\alpha}_{e\mu}F^{\alpha}_{\mu e}})\log_{2}[\frac{1}{2}(F^{\alpha}_{ee}+F^{\alpha}_{\mu\mu}-\sqrt{(F^{\alpha}_{ee}-F^{\alpha}_{\mu\mu})^{2}+4F^{\alpha}_{e\mu}F^{\alpha}_{\mu e}})]\\ \nonumber
    &+&\frac{1}{2}(F^{\alpha}_{ee}+F^{\alpha}_{\mu\mu}+\sqrt{(F^{\alpha}_{ee}-F^{\alpha}_{\mu\mu})^{2}+4F^{\alpha}_{e\mu}F^{\alpha}_{\mu e}})\log_{2}[\frac{1}{2}(F^{\alpha}_{ee}+F^{\alpha}_{\mu\mu}+\sqrt{(F^{\alpha}_{ee}-F^{\alpha}_{\mu\mu})^{2}+4F^{\alpha}_{e\mu}F^{\alpha}_{\mu e}})]\\ \nonumber
    &+&\frac{1}{2}(F^{\alpha}_{ee}+F^{\alpha}_{\tau\tau}-\sqrt{(F^{\alpha}_{ee}-F^{\alpha}_{\tau\tau})^{2}+4F^{\alpha}_{e\tau}F^{\alpha}_{\tau e}})\log_{2}[\frac{1}{2}(F^{\alpha}_{ee}+F^{\alpha}_{\tau\tau}-\sqrt{(F^{\alpha}_{ee}-F^{\alpha}_{\tau\tau})^{2}+4F^{\alpha}_{e\tau}F^{\alpha}_{\tau e}})]\\ 
    &+&\frac{1}{2}(F^{\alpha}_{ee}+F^{\alpha}_{\tau\tau}+\sqrt{(F^{\alpha}_{ee}-F^{\alpha}_{\tau\tau})^{2}+4F^{\alpha}_{e\tau}F^{\alpha}_{\tau e}})\log_{2}[\frac{1}{2}(F^{\alpha}_{ee}+F^{\alpha}_{\tau\tau}+\sqrt{(F^{\alpha}_{ee}-F^{\alpha}_{\tau\tau})^{2}+4F^{\alpha}_{e\tau}F^{\alpha}_{\tau e}})].
\end{eqnarray}
We should nevertheless note that the relative entropy of coherence for the mono-partite subsystem $e$ is equal to zero.
Similar expressions have been obtained for subsystems $\mu$ and $\tau$.

\begin{itemize}
  \item Explicit expression of genuine quantum discord in Eq.(\ref{53}):
    \end{itemize}
\begin{equation}
\begin{split}
QD^{G}(\rho^{\alpha}_{e\mu\tau})&=\frac{1}{3}[F^{\alpha}_{ee}\log_{2}F^{\alpha}_{ee}+F^{\alpha}_{\mu\mu}\log_{2}F^{\alpha}_{\mu\mu}+F^{\alpha}_{\tau\tau}\log_{2}F^{\alpha}_{\tau\tau}-(F^{\alpha}_{ee}+F^{\alpha}_{\mu\mu})\log_{2}(F^{\alpha}_{ee}+F^{\alpha}_{\mu\mu})\\
&-(F^{\alpha}_{ee}+F^{\alpha}_{\tau\tau})\log_{2}(F^{\alpha}_{ee}+F^{\alpha}_{\tau\tau})-(F^{\alpha}_{\mu\mu}+F^{\alpha}_{\tau\tau})\log_{2}(F^{\alpha}_{\mu\mu}+F^{\alpha}_{\tau\tau})\\
&+\frac{1}{2}(F^{\alpha}_{ee}+F^{\alpha}_{\mu\mu}-\sqrt{(F^{\alpha}_{ee}-F^{\alpha}_{\mu\mu})^{2}+4F^{\alpha}_{e\mu}F^{\alpha}_{\mu e}})\log_{2}[\frac{1}{2}(F^{\alpha}_{ee}+F^{\alpha}_{\mu\mu}-\sqrt{(F^{\alpha}_{ee}-F^{\alpha}_{\mu\mu})^{2}+4F^{\alpha}_{e\mu}F^{\alpha}_{\mu e}})\\
&+\frac{1}{2}(F^{\alpha}_{ee}+F^{\alpha}_{\tau\tau}-\sqrt{(F^{\alpha}_{ee}-F^{\alpha}_{\tau\tau})^{2}+4F^{\alpha}_{e\tau}F^{\alpha}_{\tau e}})\log_{2}[\frac{1}{2}(F^{\alpha}_{ee}+F^{\alpha}_{\tau\tau}-\sqrt{(F^{\alpha}_{ee}-F^{\alpha}_{\tau\tau})^{2}+4F^{\alpha}_{e\tau}F^{\alpha}_{\tau e}})\\
 &+\frac{1}{2}(F^{\alpha}_{ee}+F^{\alpha}_{\tau\tau}-\sqrt{(F^{\alpha}_{ee}-F^{\alpha}_{\tau\tau})^{2}+4F^{\alpha}_{e\tau}F^{\alpha}_{\tau e}})\log_{2}[\frac{1}{2}(F^{\alpha}_{ee}+F^{\alpha}_{\tau\tau}-\sqrt{(F^{\alpha}_{ee}-F^{\alpha}_{\tau\tau})^{2}+4F^{\alpha}_{e\tau}F^{\alpha}_{\tau e}})]\\ 
    &+\frac{1}{2}(F^{\alpha}_{ee}+F^{\alpha}_{\tau\tau}+\sqrt{(F^{\alpha}_{ee}-F^{\alpha}_{\tau\tau})^{2}+4F^{\alpha}_{e\tau}F^{\alpha}_{\tau e}})\log_{2}[\frac{1}{2}(F^{\alpha}_{ee}+F^{\alpha}_{\tau\tau}+\sqrt{(F^{\alpha}_{ee}-F^{\alpha}_{\tau\tau})^{2}+4F^{\alpha}_{e\tau}F^{\alpha}_{\tau e}})\\
     &+\frac{1}{2}(F^{\alpha}_{\mu\mu}+F^{\alpha}_{\tau\tau}-\sqrt{(F^{\alpha}_{\mu\mu}-F^{\alpha}_{\tau\tau})^{2}+4F^{\alpha}_{\mu\tau}F^{\alpha}_{\tau \mu}})\log_{2}[\frac{1}{2}(F^{\alpha}_{\mu\mu}+F^{\alpha}_{\tau\tau}-\sqrt{(F^{\alpha}_{\mu\mu}-F^{\alpha}_{\tau\tau})^{2}+4F^{\alpha}_{\mu\tau}F^{\alpha}_{\tau \mu}})]\\ 
    &+\frac{1}{2}(F^{\alpha}_{\mu\mu}+F^{\alpha}_{\tau\tau}+\sqrt{(F^{\alpha}_{\mu\mu}-F^{\alpha}_{\tau\tau})^{2}+4F^{\alpha}_{\mu\tau}F^{\alpha}_{\tau \mu}})\log_{2}[\frac{1}{2}(F^{\alpha}_{\mu\mu}+F^{\alpha}_{\tau\tau}+\sqrt{(F^{\alpha}_{\mu\mu}-F^{\alpha}_{\tau\tau})^{2}+4F^{\alpha}_{\mu\tau}F^{\alpha}_{\tau \mu}})].
\end{split}
\label{54}
\end{equation}

\end{document}